\documentclass[usenatbib]{mn2e}
\usepackage{amssymb}
\usepackage{graphicx}
\newcommand{\text}[1]{\quad\mbox{#1}\quad}

\newcommand{\oder}[2]{\frac{d #1}{d #2}}

\newcommand{\by}{\!\times\!}

\newcommand{\azh}{Astron. Zh.}
\newcommand{\apj}{ApJ}
\newcommand{\apjs}{ApJS}
\newcommand{\apjl}{ApJ}
\newcommand{\mnras}{MNRAS}

\newcommand{\nat}{Nature}
\newcommand{\araa}{ARA\&A}

\newcommand{\physrep}{Phys. Report}
\newcommand{\prd}{Phys. Rev. D}
\newcommand{\iaucirc}{IAU Circ.}
\newcommand{\fracb}[2]{\left(\frac{#1}{#2}\right)}
\newcommand{\mean}[1]{\langle{#1}\rangle}
\newcommand{\sub}[1]{_{\mbox{\tiny #1}}}
\newcommand{\sups}[1]{^{\mbox{\tiny #1}}}

\title[Recycling of Neutron Stars and Hypernovae]
{Recycling of Neutron Stars in Common Envelopes and Hypernova Explosions} 

\author[M.V.~Barkov and S.S. Komissarov]
{Maxim V.~Barkov,$^{1,2,3}$\thanks{E-mail: bmv@maths.leeds.ac.uk} and 
Serguei S.~Komissarov$^{2}$\thanks{E-mail: serguei@maths.leeds.ac.uk}\\ \\
$^{1}$Max-Planck-Institut f\"ur Kernphysik, Saupfercheckweg 1, 69117 Heidelberg, Germany\\
$^{2}$Department of Applied Mathematics, The University of Leeds,
Leeds, LS2 9JT, UK\\
$^{3}$Space Research Institute, 84/32 Profsoyuznaya Street, Moscow
117997, Russia}

\begin{document}
\date{Received/Accepted}
\maketitle
                                                                              
\begin{abstract}

In this paper we propose a new plausible mechanism of supernova explosions 
specific to close binary systems. 
The starting point is the common envelope phase in the evolution 
of a binary consisting of a red super giant and a neutron star.   
As the neutron star spirals towards the center of its companion it 
spins up via disk accretion. Depending on the specific angular momentum 
of gas captured by the neutron star via the Bondi-Hoyle mechanism, 
it may reach  millisecond periods either when it is still inside the 
common envelope or after it has merged with the companion core. 
The high accretion rate may result in strong differential rotation of 
the neutron star and generation of the magnetar-strength magnetic field. 
The magnetar wind can blow away the common envelope if its  
magnetic field is as strong as $10^{15}\,$G, and can destroy the entire 
companion if it is as strong as $10^{16}\,$G. The total explosion energy can be 
comparable to the rotational energy of a millisecond pulsar and reach  
$10^{52}\,$erg. However, only a small amount of $^{56}$Ni is expected to 
be produced this way. The result is an unusual type-II supernova with very high 
luminosity during the plateau phase, followed by a sharp drop in brightness 
and a steep light-curve tail. The remnant is either a solitary magnetar or a  
close binary involving a Wolf-Rayet star and a magnetar. When this Wolf-Rayet 
star explodes this will be a third supernovae explosion in the same binary.   
\end{abstract}

\section{Introduction}
\label{intro}

Usually, neutron stars (NS) have magnetic field $B \sim 10^{12}$G and 
rotate with a period of fraction of a second, the Crab pulsar being a
typical example. However, we now know that the ``zoo'' of NS is much more 
diverse and they can have both much weaker and much stronger magnetic field 
and rotate with both much longer and much shorter periods. 

Around 10\% of NS have surface dipolar magnetic field $B\sim 10^{14}-10^{15}$G  
\citep{kd98}. These ``magnetars'' are believed to be born in core collapse 
explosions of rapidly rotating stars. During the collapse, the proto-neutron star 
naturally develops strong differential rotation, which allows for generation 
of magnetic field via $\alpha$-$\Omega$-dynamo \citep{DT92,TD93}. 
The strength of saturated magnetic field strongly depends on the rotational
period, with shorter periods leading to stronger magnetic field and more rapid 
release of rotational energy. In order to generate the magnetar strength 
magnetic field the rotation period must be around few milliseconds \citep{DT92}. 
Such a strong field allows to release up to $10^{52}\,$erg of magnetar's 
rotational energy in very short period of time. This is sufficient to 
drive extremely powerful supernova explosions, on the hypernova scale,   
and to produce powerful Gamma Ray Bursts \citep[GRB, e.g.][]{U92,tcq04}.  
Turbulence required for the magnetic dynamo action can also be generated via
the magneto-rotational instability \citep[MRI,][]{BH91}. Calculations
based on the linear theory show that in the supernova context strong
saturation field can be reached very quickly, on the time scale of only several
tens of rotational periods \citep{AWML03,ocma09}. 

Millisecond pulsars are found in low mass binaries, and it is generally thought that 
they have been spun up via disk accretion \citep{acrs82,ar09}. 
This origin implies mass increase by about 0.2$M_\odot$ compared to normal 
radio pulsars, whose masses are narrowly distributed around $1.35M_\odot$ 
\citep{ts99}. It is rather difficult to measure the mass of a millisecond pulsar, but 
the few available results agree with this prediction of the accretion model 
\citep{ktr94,jh05,dp10}. The most massive pulsar found to date, almost 2$M_\odot$, 
is a millisecond pulsar \citep{dp10}. 
The magnetic field of these millisecond pulsars  is very low, down to $10^9$~Gauss. 
Most likely, their initial magnetic field was of similar strength to 
normal pulsars, but now it is buried under the layers of accreted matter 
\citep{bkk74,acrs82}.   
The reason why these pulsars could not generate magnetar-strength magnetic field in 
the same way as in the core-collapse scenario is the very long time scale of 
spinning up compared  to the viscous time-scale. As the result, the differential 
rotation remains weak and there is not enough energy for effective magnetic 
dynamo \citep{s99}. 

The rotational frequency of the fastest X-ray pulsar is $\sim760$~Hz \citep{cmm03}, 
whereas the fastest known radio pulsar has the frequency of $641$~Hz \citep{bkh82}.
Most likely, it is the gravitational radiation losses what places the upper limit on the
rotation rates because at high spin the r-mode oscillations become 
excited \citep{ST83,l99}. \citet{s99} argued that this instability may also 
result in magnetic explosion. 
The idea is that the heating of NS, associated with these oscillations, 
reduces its viscosity and decouples its interior from the outer layers. 
Being most disturbed, the outer layers rapidly loose some of their angular 
momentum via gravitational radiation. This leads to strong differential rotation 
and generation of magnetar strength magnetic field in the NS interior. 
This field becomes unstable to buoyancy, emerges on the surface, 
and magnetically driven pulsar wind rapidly extracts  the rotational energy of 
the NS. \citet{s99} proposed this as an alternative scenario for long GRBs. 
It is unlikely that a supernova-like event can accompany a GRB in this scenario. 
Although the wind energetics is sufficient, only a small fraction of this 
energy can be deposited into the companion star, simply because of its small 
geometrical cross-section. Moreover, recent results suggest that the amplitude of r-modes 
may saturate at a much lower level due to nonlinear interaction with other modes 
\citep{afm03,btw05,btw07}.

A similar recycling of NS may occur during the common envelop (CE) phase, 
after the primary becomes a red super giant \citep[RSG; ][]{bks71,P76,ty79,PY06}. Due to the 
dynamic friction, the NS then spirals inside the RSG, accreting on its way. 
Now one can imagine two interesting outcomes of such process.  
First, the neutron star may accumulate too much mass and collapse into a 
black hole (BH). This BH is likely to be rapidly rotating and  
drive a stellar explosion in the collapsar fashion \citep{fw98,zf01,BK10}. 

Second, the NS may first spin up to a millisecond period and drive a  
magnetic explosion of the type proposed by \citet{s99} but now inside the 
common envelope. The magnetar wind can keep energising such supernovae, 
producing a similar effect to radioactive decay \citep{w10,kb10}. High accretion 
rates may modify the way the NS is recycled. The accreted gas can form a massive 
rapidly rotating layer above the NS crust \citep{is99,is10}. The strong differential 
rotation between the layer and the NS core may result in development of the 
Kelvin-Helmholtz instability when the NS crust melts down under the weight of 
the layer. This may lead to turbulence and strong amplification of the NS 
magnetic field.  

The accretion onto neutron stars during the in-spiral has been studied by \citet{ch96}, 
who concluded that the high angular momentum of the gas gravitationally captured 
by NS prevents it from effective neutrino cooling and keeps the mass accreting 
rate well below the rate of the Bondi-Hoyle capture. As the result, the NS 
accumulates very little mass while still inside the common envelope. On the 
other hand, he suggested that during the merger with the companion core the 
mass accretion rate rises and the NS collapses into a black hole. While carefully
analysing various effects of rotation, \citet{ch96} did not use any particular 
model for the primary. Moreover, there is a great deal of uncertainty with the 
regard to the specific angular momentum of the gravitationally captured gas. 
In our study we come back to this problem, consider realistic models of RSG 
stars and allow for the uncertainty. 

The paper is organised as follows. In Section 2 we consider the accretion and 
recycling of NSs during the in-spiral. We conclude that the outcome is very
sensitive to the assumed specific angular momentum of the gravitationally 
captured gas. Given the current uncertainty with the regard to the angular 
momentum, it seems possible that NSs begin to accrete with Bondi-Hoyle rate 
while still inside common envelopes. In this case, they rapidly spin up to 
millisecond periods.  
In Section 3 we speculate on the possible mechanisms of generating magnetar-strength 
magnetic field  and study the magnetic interaction of millisecond magnetars with 
accretion flows typical for the in-spiral problem. 
We find that if the magnetar forms inside the common 
envelope and its magnetic field is about $10^{15}\,$G, the magnetospheric 
pressure can overcome the gravity and drive an outflow, with eventual release 
of up to $10^{52}$erg of the magnetar rotational energy. If the magnetar forms only 
after the merger with the core then higher magnetic field, $\sim10^{16}$G, is 
required to drive stellar explosions. In Section 4 we discuss the properties of such 
explosions and their observational signatures.  Because the RSG envelope is rich 
in hydrogen the supernova will be classified as type-II. Because of the very 
high energy release and relatively small mass of the RSG, the speed of the ejecta 
is expected to be very high, $\sim 10^9$cm/s, and the luminosity at the 
plateau phase $\sim 10^{43}$erg/s. However, due to the small amount of 
$^{56}$Ni generated in the explosion and the rapid spin-down of the magnetar, the 
plateau is followed by a sharp drop in brightness and a steep light-curve tail. 
We show that the termination shock of the magnetar wind produces a high energy 
synchrotron and inverse Compton emission which may energise the tail, but 
the supernova ejecta soon becomes transparent to this emission, limiting its 
potential to mimic the effect of radioactive decay. 
The flux of gamma-ray emission is expected to be rather low and difficult to 
observe, unless the explosion occurs in the Local Group of galaxies. 
In the case of off-center explosion, the remnant is a very close 
binary system consisting of a WR star and a magnetar, but the strong magnetic 
field prevents magnetar from accreting plasma of the WR wind. 
Our results are summarized in Section 5.           
   
In this paper, the dimensional estimates are presented using the 
following notation: 
the time $t\sub{x,n}$ is measured in $10^n$s,
the distance $R\sub{x,n}$  in $10^n\,$cm, 
the speed $V\sub{x,n}$ in $10^n\,\mbox{cm}/\mbox{s}$, 
the mass $M\sub{x,n}$ in $10^nM_\odot$, 
mass accretion rate $\dot{M}\sub{x,n}$ in $10^n M_\odot/\mbox{yr}$,
and the magnetic field $B\sub{x,n}$ in $10^n$G.

\section{In-spiral dynamics and recycling of neutron stars}
\label{In-spiral}

As the neutron spirals inside its giant companion it accretes mass and 
angular momentum. The accretion can proceed at the Eddington rate or at 
the much higher Bondi-Hoyle rate. This depends on whether the radiation 
is trapped in the accretion flow and the neutrino cooling is sufficiently 
effective \citep{hch91,ch93,ch96,fbh96}.  
Only if the accretion proceeds at the Bondi-Hoyle rate the recycling of NS 
is sufficiently fast and the explosion can occur during the inspiral.  
  
In the Bondi-Hoyle problem, an accreting 
point mass $M$ is moving with finite speed $v_\infty$ through a uniform medium 
of mass density $\rho_\infty$ and sound speed $c_{\rm s,\infty}$.  The Bondi-Hoyle 
mass accretion rate is well approximated by the equation   
\begin{equation}
\dot{M}\sub{BH} = \pi R\sub{A}^2\rho_\infty v_\infty,
\label{bondi-hoyle}
\end{equation}
where
\begin{equation}
R\sub{A}=\delta({\cal M}_\infty)\frac{2GM}{v_\infty^2}
\label{ra1}
\end{equation}
is the accretion radius, and   
\begin{equation}
\delta({\cal M}_\infty)=\fracb{{\cal M}_\infty^2}{1+{\cal M}_\infty^2}^{3/4},
\end{equation}
where ${\cal M}_\infty = v_\infty/c_{\rm s,\infty}$ is the Mach number~\citep{smts85}.
The mean angular momentum of accreted matter is zero. 

The in-spiral problem is more complicated due to the density and 
velocity gradients across the direction of motion of the NS.  
Provided, the gradients are small  
on the scale of accretion radius the above expression for $\dot{M}$ is still 
quite accurate \citep{im93,r97,r99}.
However, the accreted matter can now have non-vanishing mean angular momentum.  
This was first suggested by \citet{is75} in connection with  
X-ray binaries, where the accretion disk of the black hole is fed by the stellar wind of 
the primary. Taking into account only the velocity gradients related to the orbital 
motion, they found that the mean specific angular momentum of accreted matter inside 
the accretion cylinder is 
\begin{equation}
\mean{j\sub{A}}=\frac{1}{4}\Omega R\sub{A}^2,
\label{ja1}
\end{equation}
where $\Omega$ is the angular velocity of the orbital motion. 
\citet{sl76} took into account the density gradient of the wind as well 
and refined this result, 
\begin{equation}
\mean{j\sub{A}}=-\frac{1}{2}\Omega R\sub{A}^2,
\label{ja2}
\end{equation}
where the sign minus shows that the disk rotation is now in the opposite sense to 
the orbital motion. Later, \citet{dp80} criticised the approach used by the previous 
authors. They generalised the original calculations of \citet{hl39} to include the 
density and velocity gradients and concluded that to the first order in $R\sub{A}/h$, 
where $h$ is the length scale of the variations, $\mean{j\sub{A}}=0$. 
The 2D numerical simulations of \citet{im93} 
seemed to agree with this conclusion, whereas their 3D simulations still indicated 
net accretion of angular momentum. 
\citet{r97,r99} carried out extensive 3D numerical investigation 
of the problem. Introducing angle-dependent accretion radius, 
he also derived another version of the analytic expression for  
$\mean{j\sub{A}}$ 
\begin{equation}
\mean{j\sub{A}}=\frac{1}{4}(6\epsilon\sub{v}+\epsilon_\rho)v_0 R\sub{A},
\label{ja3}
\end{equation}
where 
\begin{equation}
\epsilon_\rho=\frac{R\sub{A}}{\rho}\oder{\rho}{x},\qquad 
\epsilon_\rho=\frac{R\sub{A}}{v}\oder{v}{x}
\label{epsi}
\end{equation}
are the dimensionless gradients of density and velocity respectively, 
with $x$ being the transverse Cartesian coordinate. However, the numerical 
results did not agree with this expression, with $\mean{j\sub{A}}$ varying between 
7\% and 70\% of what is available in the accretion cylinder. 
Moreover, in the simulations only 
the cases with either velocity or density gradient were considered, whereas in 
the in-spiral problem both are present. Although the computational resources and 
numerical algorithms have improved since 1999,  no further attack on this problem 
has been attempted yet and the issue of $\mean{j\sub{A}}$ remains open. 

For the inspiral problem $v_0=\Omega a$, $v(x)=v_0+\Omega x$, 
and $\epsilon\sub{v}=R\sub{A}/a$, where $a$ is the orbital radius of NS. 
Assuming power law for density distribution inside the RSG, $\rho\propto r^{-n}$, 
we also find $\epsilon_\rho=-nR\sub{A}/a$. This allows us to write Eq.\ref{ja3} as 
\begin{equation}
\mean{j\sub{A}}=\frac{1}{4}(6-n) \Omega R\sub{A}^2,
\label{ja5}
\end{equation}
which has the same dependence on $\Omega$ and $R\sub{A}$ as in Eqs.\ref{ja1} and 
\ref{ja2}. Given the unsettled nature of this issue, we will assume that 
\begin{equation}
\mean{j\sub{A}}=\frac{\eta}{4}\Omega R\sub{A}^2,
\label{ja}
\end{equation}
where $\eta$ is a free parameter, which reflects our current ignorance. 

In our case, $M$ is the mass of the NS and $v_\infty$ is its  
orbital speed. Assuming that the interaction does not disturb the RSG inside
the orbit, we have  
\begin{equation}
 v^2 \simeq G M_*/a,
\label{os}
\end{equation}
where $a$ is the orbital radius and $M_*$ is the mass of RSG inside the 
orbit. Under the same assumption,  
$\rho_\infty$ and $c_{\rm s,\infty}$ can be replaced with the density 
and sound speed of the undisturbed RSG at radius $r=a$.   
The motion of the NS inside RSG is only mildly supersonic,  
${\cal M} \simeq 1.4$ \citep{ch93}, allowing us to write 
\begin{equation} 
R\sub{A} \approx 2\beta \frac{aM\sub{NS}}{M_*},
\label{ras}
\end{equation} 
where $\beta\simeq0.8$ and only weakly depends on the model of RSG.

Given the specific angular momentum we can estimate the distance from the NS
at which the Bondi-Hoyle trapped gas will form an accretion disk as 
\begin{equation} 
R\sub{c} = \frac{\mean{j\sub{A}}^2}{GM\sub{NS}} \approx
a \eta^2 \beta^4 \fracb{M\sub{NS}}{M_*}^3.
\label{rc}
\end{equation}
In the literature this radius is often called the circularisation radius. 
Since the NS is not a point mass the accretion disk forms only if $R\sub{c}>R\sub{NS}$. 
When this condition is satisfied, the recycling of NS proceeds at the highest rate. 

For $R\gg R_c$ the accretion on NS may proceed in more or less spherical fashion and
is well described by    
the Bondi solution, in which the accretion flow becomes supersonic
at $R\sub{acc}=0.25R\sub{A}$ (for the polytropic index $\gamma=4/3$).  
Its collision with the NS and/or its accretion disk creates a shock wave which 
re-heats the flow. This will have no effect on the mass accretion rate only if  
the following two conditions are satisfied \citep{ch96}.   

Firstly, the radiation produced by the gas heated at the accretion shock should 
not be able to escape beyond $R\sub{A}$. If it does the mass accretion rate is 
limited by the Eddington value
$$
\dot{M}\sub{Edd} \sim 3\by 10^{-8} M_{\odot}\,\mbox{yr}^{-1}. 
$$
This condition is satisfied when the accretion shock radius
\begin{equation}
R\sub{sh}\approx 3\by 10^9 R\sub{c,6}^{1.48}
\dot{M}\sub{0}^{-0.37}\,\mbox{cm}
\label{rsh}
\end{equation}
is smaller compared to the radiation trapping radius
\begin{equation}
R\sub{tr}\approx 3.4\by 10^{13}
\dot{M}\sub{0}\, \mbox{cm} . 
\label{rtr}
\end{equation}
For this to occur $\dot{M}\sub{BH}$ 
has to exceed the critical value 
\begin{equation}
\dot{M}\sub{cn1} \approx 1.1 \by 10^{-3}
R\sub{c,6}^{1.08} 
M_{\odot} \mbox{ yr}^{-1}
\label{dmcn1}
\end{equation}
\citep{ch96}. This result was derived assuming that $R\sub{c}>R\sub{NS}$ and ignoring 
the General Relativity corrections. 

Secondly, the shock has to be inside the sonic surface of the Bondi 
flow\footnote{The shock radius in Eq.\ref{rsh} is obtained assuming cold supersonic 
flow at infinity}. Otherwise, this shock cannot be a part of the stationary 
solution and hence the Bondi-type accretion cannot be realised. This condition is 
satisfied when the $\dot{M}\sub{BH}$ exceeds the critical value 
\begin{equation}
 \dot{M}\sub{cn2} \approx  10^{4}
 R\sub{acc,8}^{-2.7} 
 R\sub{c,6}^{4} 
 M_{\odot}\,\mbox{yr}^{-1}
\label{dmcn2}
\end{equation}
\citep{ch96}. As we shell see this condition is more restrictive.  

A note of caution has to be made at this point. The analytical 
expression for the accretion shock radius (\ref{rsh}) is based on a number 
of assumptions and simplifications and its accuracy has not yet been tested 
via numerical simulations. At the moment, this can only be considered as an order
of magnitude estimate. When $R\sub{c}$ becomes as small as the NS radius, this 
expression gives $R\sub{sh}$ which exceeds the value given by the more reliable 
spherically symmetric model of \citet{hch91} by a factor of ten (see Eqs. 
\ref{rsh} and \ref{r-sh1}). Two reasons 
explain this disagreement. First, in the spherically symmetric models the gravity 
is relativistic, whereas in the models with rotation it is Newtonian. Second, 
the models use different assumptions on the geometry of the neutrino cooling 
region. It appears that Eq.\ref{rsh}  becomes increasingly less accurate 
when $R\sub{c}$ approaches $R\sub{NS}$, overestimating $R\sub{sh}$. As the 
result, Eq.\ref{dmcn2} overestimates the value of $\dot{M}\sub{cn2}$.
We also note, that the critical rates are quite sensitive 
to the parameter $\eta$, 
$$
   \dot{M}\sub{cn1} \propto \eta^{2.16} \text{and} \dot{M}\sub{cn2} \propto \eta^{8}.
$$   

In this study, we have considered six different models of RSG. In the first two 
models we assumed that mass distribution in the stellar envelope is the  
power law , $\rho\propto r^{-n}$, with $n=2$ (model RSG2) and $n=3$ (model RSG3). 
In both these models, the total stellar mass $M_*=25 M_{\odot}$, 
the helium core mass  $M_{c} = 1/3 M_*$, 
the inner and the outer radii of the power law envelope are 
$7\by 10^9$ cm and $7\times 10^{13}$ cm respectively. 
For these models all calculations can be made analytically. Not only this 
helps to develop a feel for the problem, but also provides useful test cases 
for numerical subroutines. 
The next two models are based on stellar evolution calculations and  
describe RSG in the middle of the He-burning phase \citep{hwls04,whw02}. 
Their Zero Age Main Sequence (AMS) masses are $M=20 M_{\odot}$ (model RSG20He) 
and $M=25 M_{\odot}$ (model RSG25He). These data were kindly provided to us by 
Alexander Heger.   
The last two models, RSG20 and RSG25, describe pre-supernova RSG with  
the same AMS masses\footnote{These models can be downloaded from the website 
http://homepages.spa.umn.edu/$\sim$alex/stellarevolution/ .}. 
Their have very extended envelopes, increasing the chance of CE phase.  

\begin{figure*}
\includegraphics[width=84mm,angle=-0]{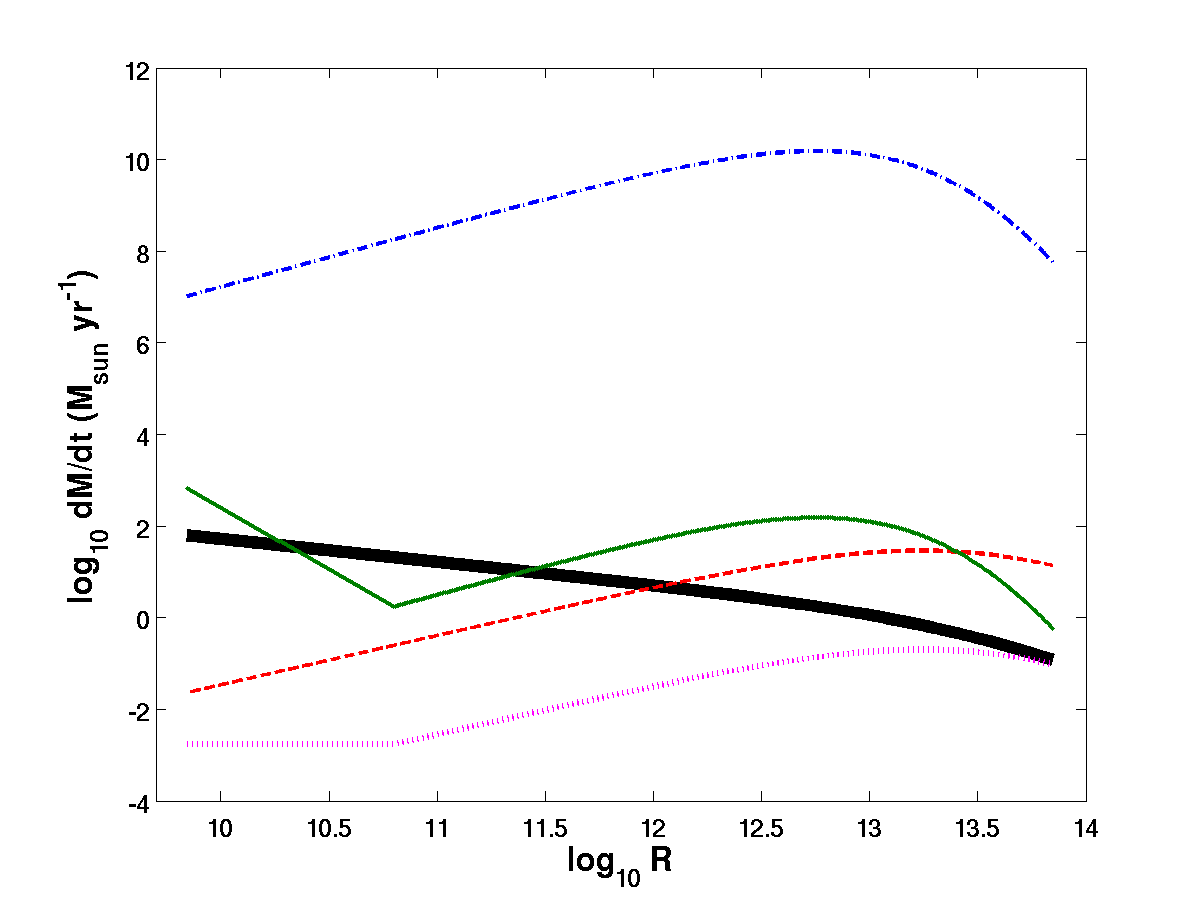}
\includegraphics[width=84mm,angle=-0]{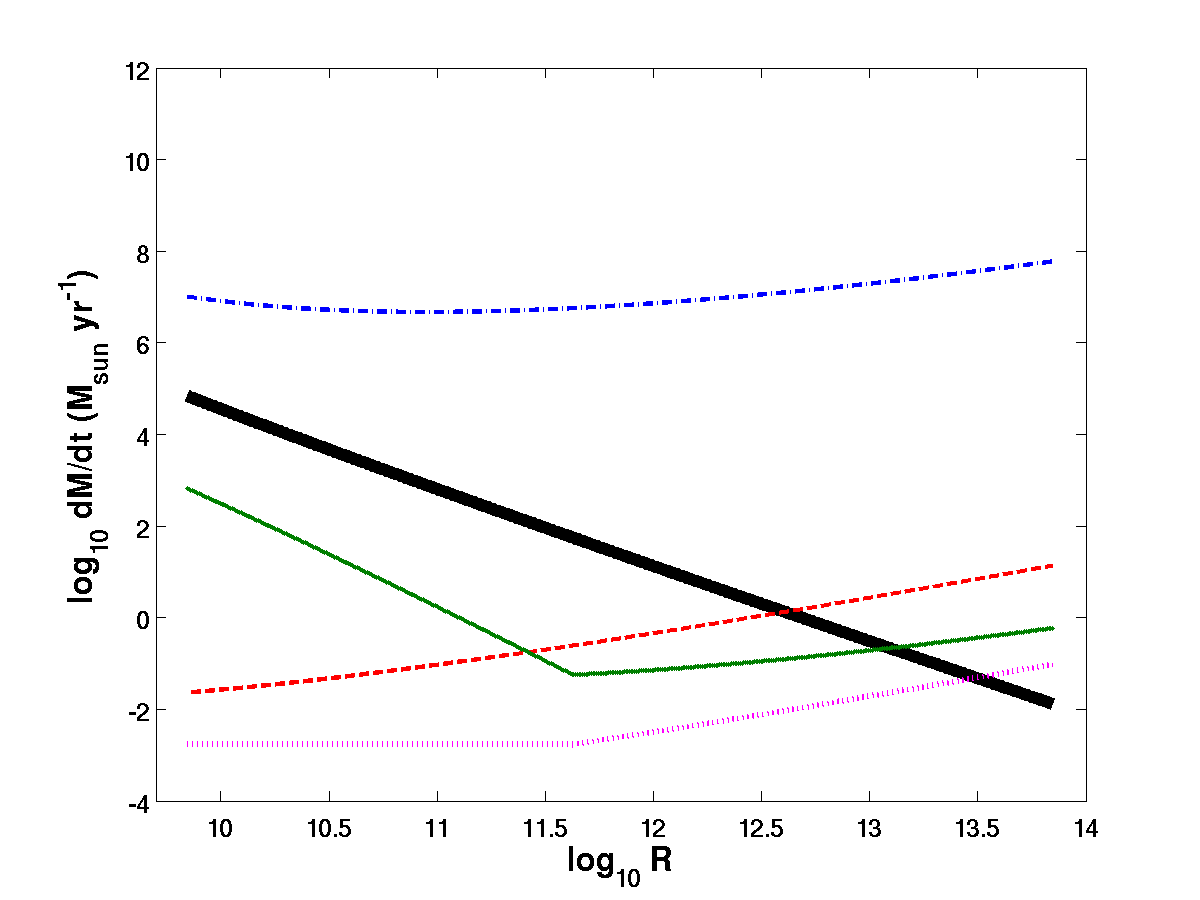}
\includegraphics[width=84mm,angle=-0]{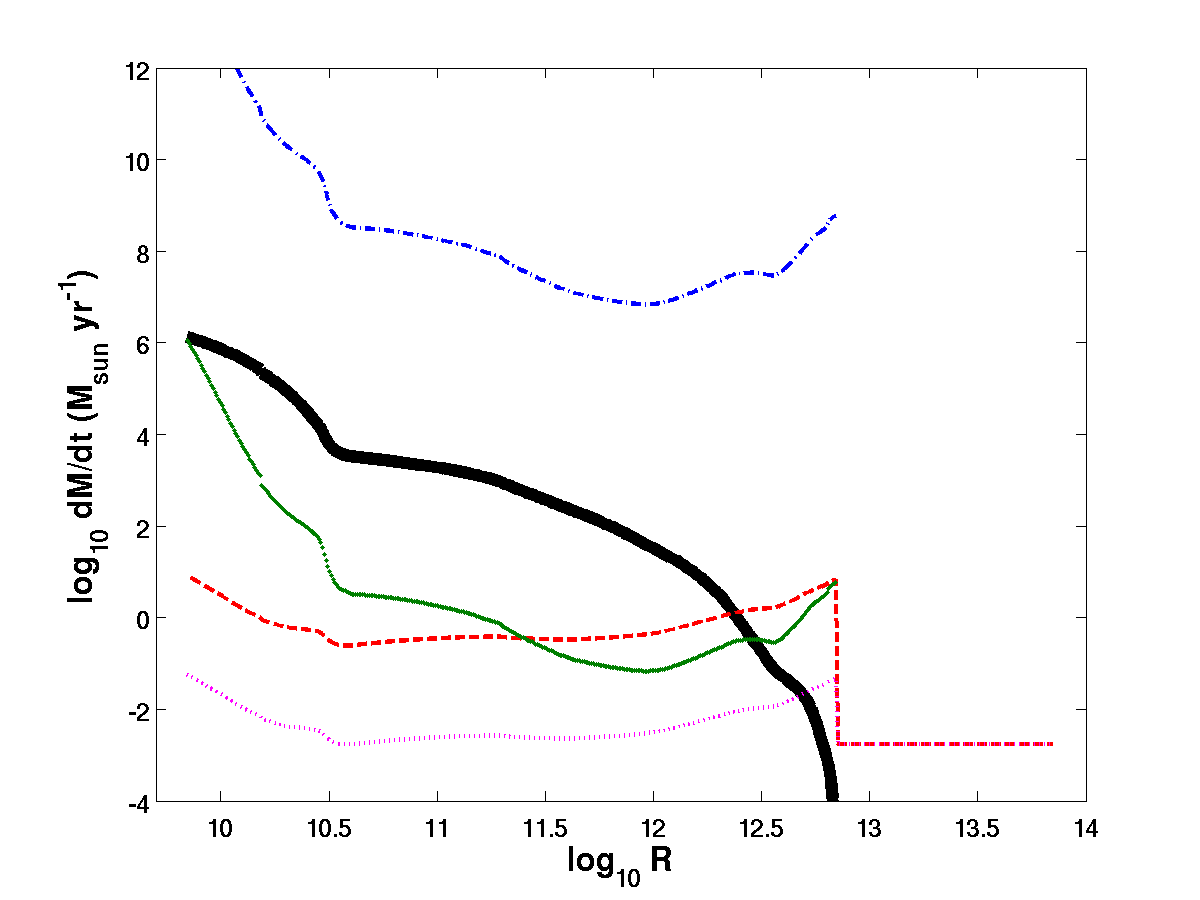}
\includegraphics[width=84mm,angle=-0]{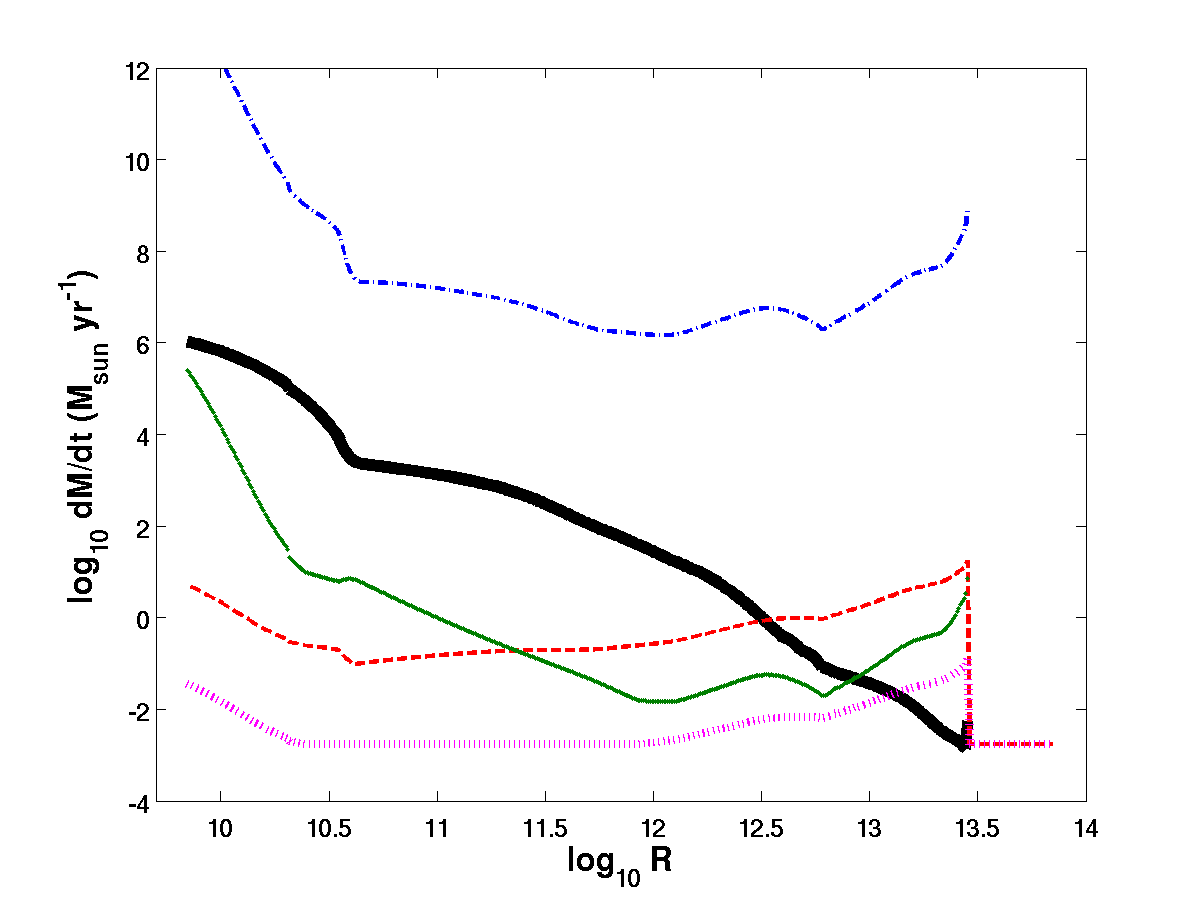}
\includegraphics[width=84mm,angle=-0]{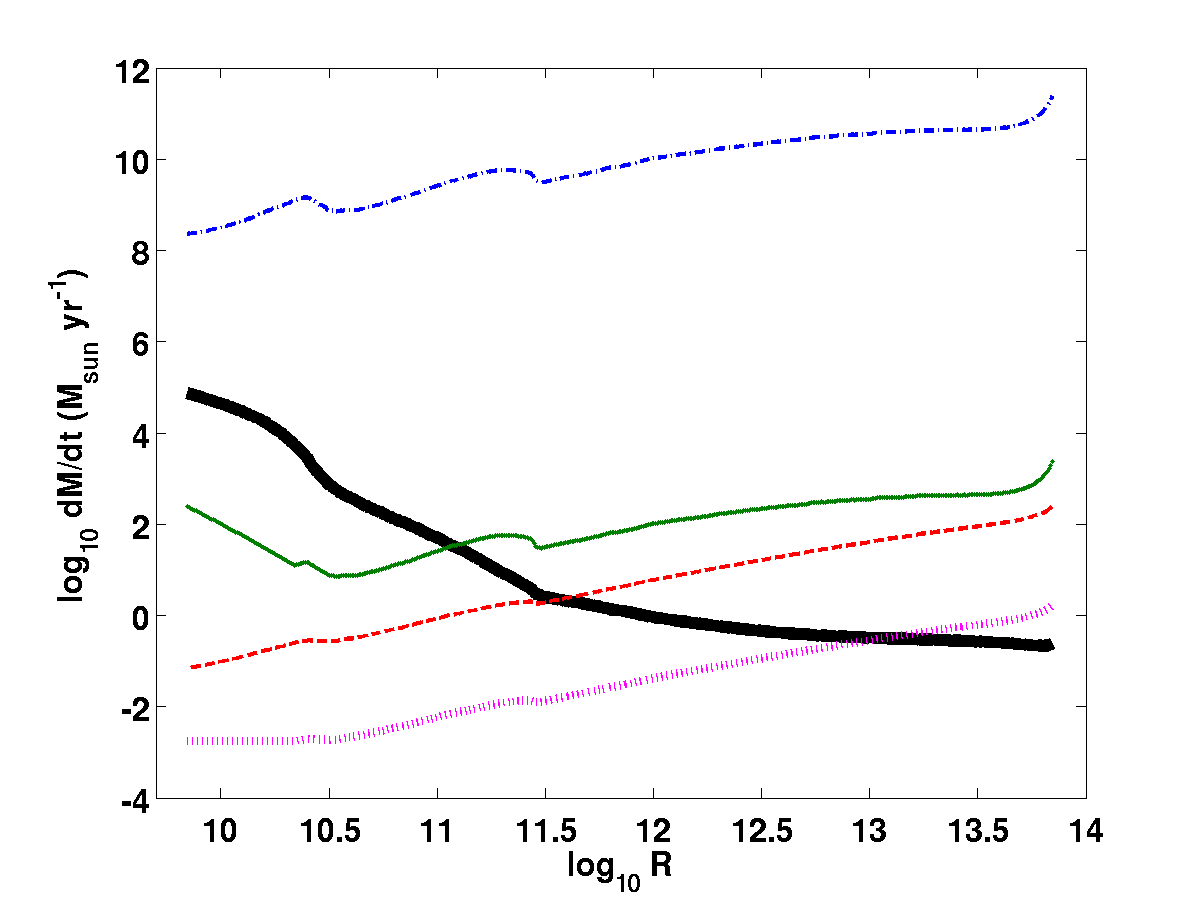}
\includegraphics[width=84mm,angle=-0]{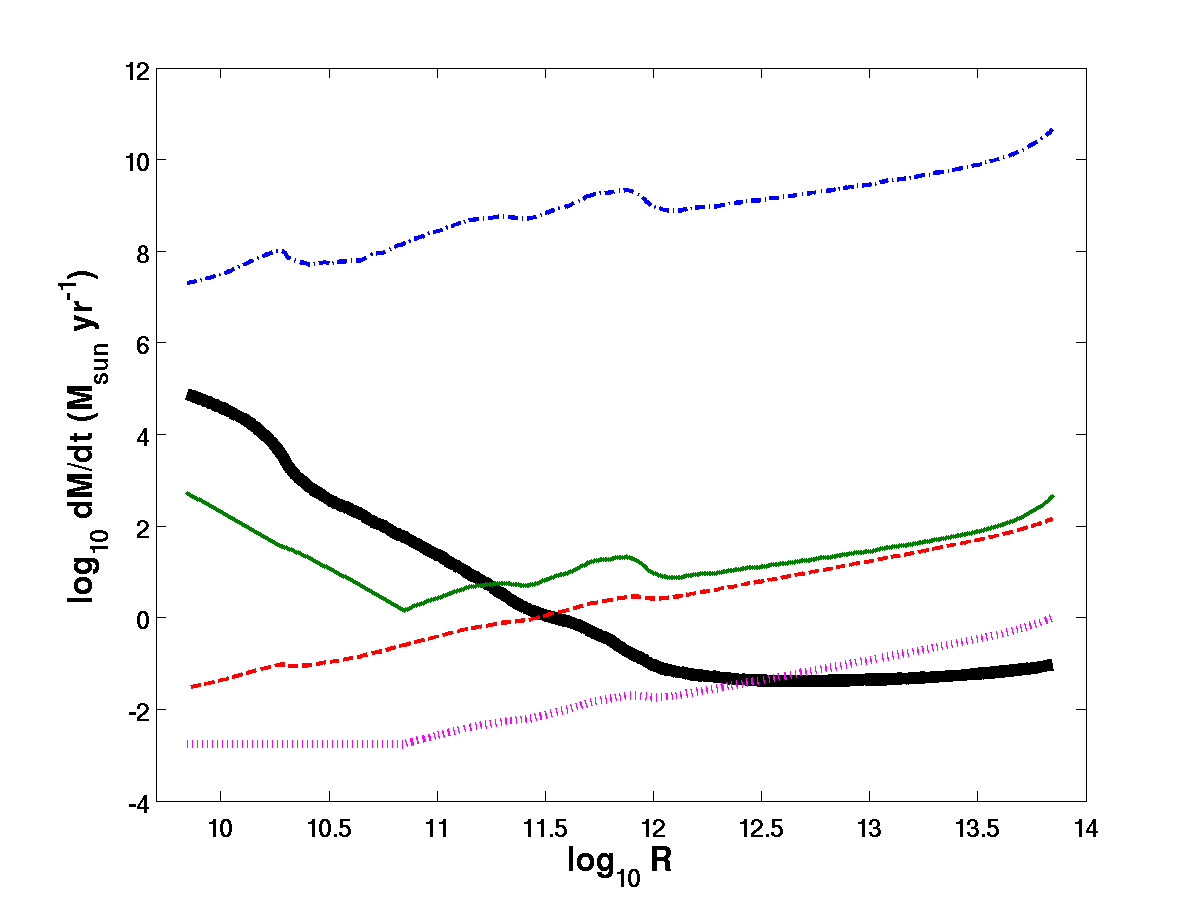}
\caption{Accretion rate as a function of position inside the 
RSG primary for six different models: 
RSG2 (left top panel), RSG3 (right top panel), RSG20He (left middle panel),  
RSG25He (right middle panel), RSG20 (left bottom panel), and RSG25 (right bottom panel).
The Bondi-Hoyle rate, $\dot{M}\sub{B-H}$, is shown by thick solid lines, 
the critical value $\dot{M}\sub{cn1}$ by dotted ($\eta=0.1$) and 
dashed ($\eta=1$) lines, 
and the critical value $\dot{M}\sub{cn2}$ by thin solid ($\eta=0.1$) and 
dot-dashed ($\eta=1$) lines. 
}
\label{dmdtf}
\end{figure*}

\begin{figure*}
\includegraphics[width=84mm,angle=-0]{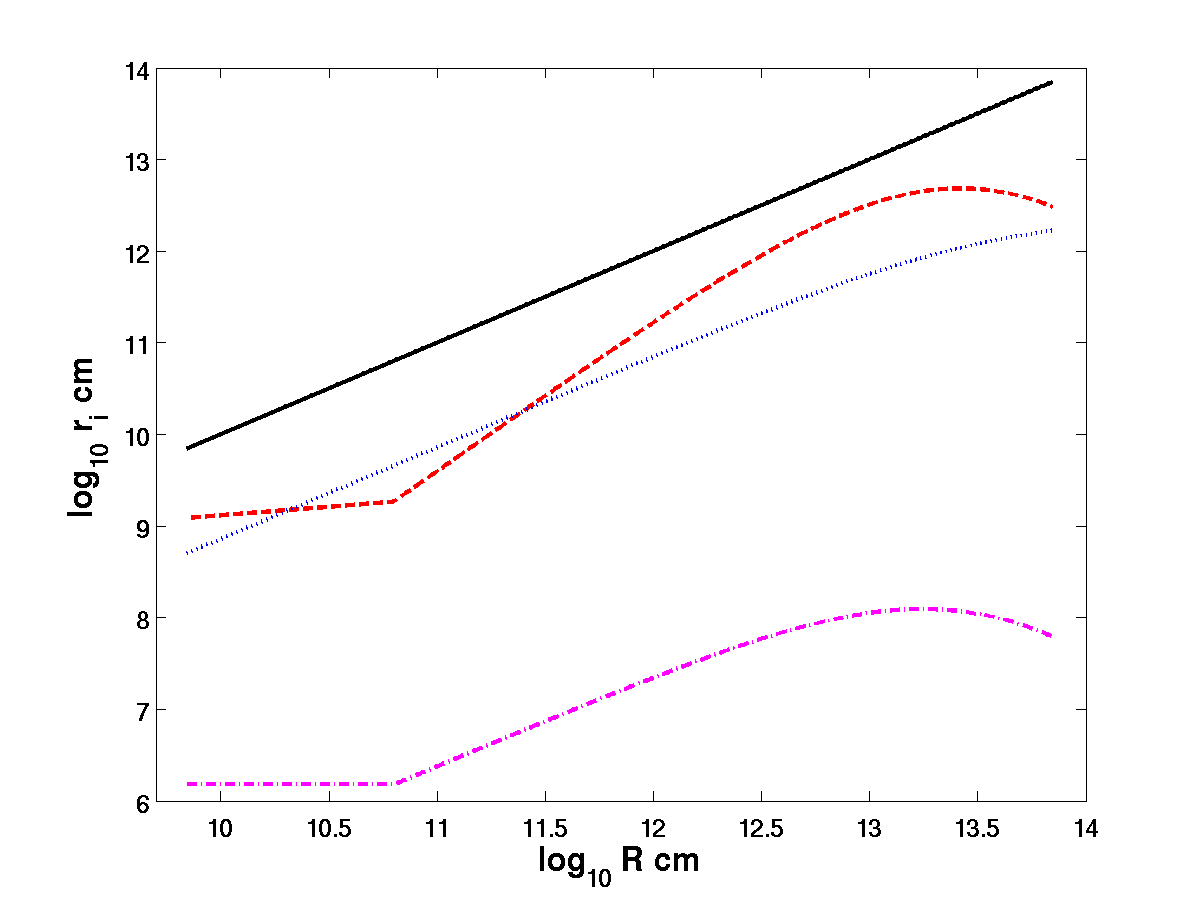}
\includegraphics[width=84mm,angle=-0]{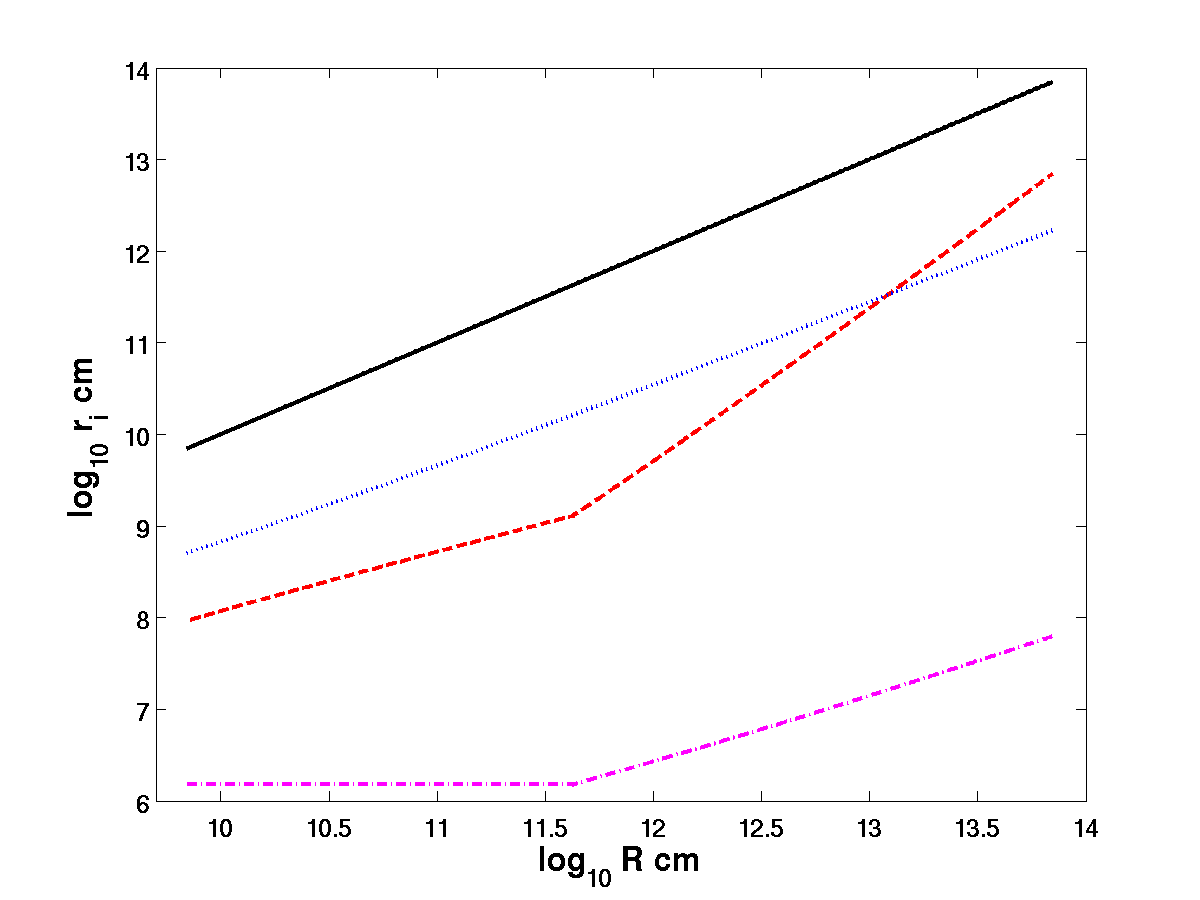}
\includegraphics[width=84mm,angle=-0]{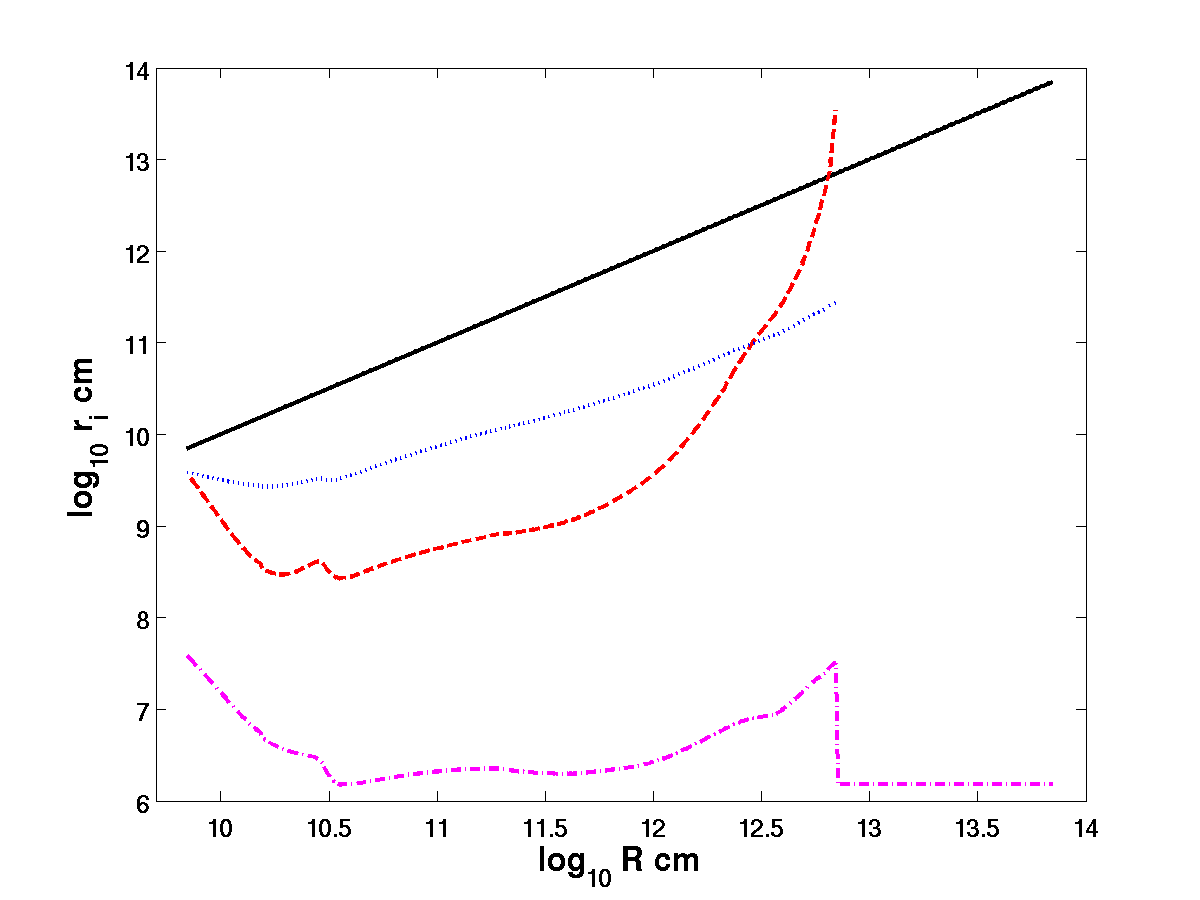}
\includegraphics[width=84mm,angle=-0]{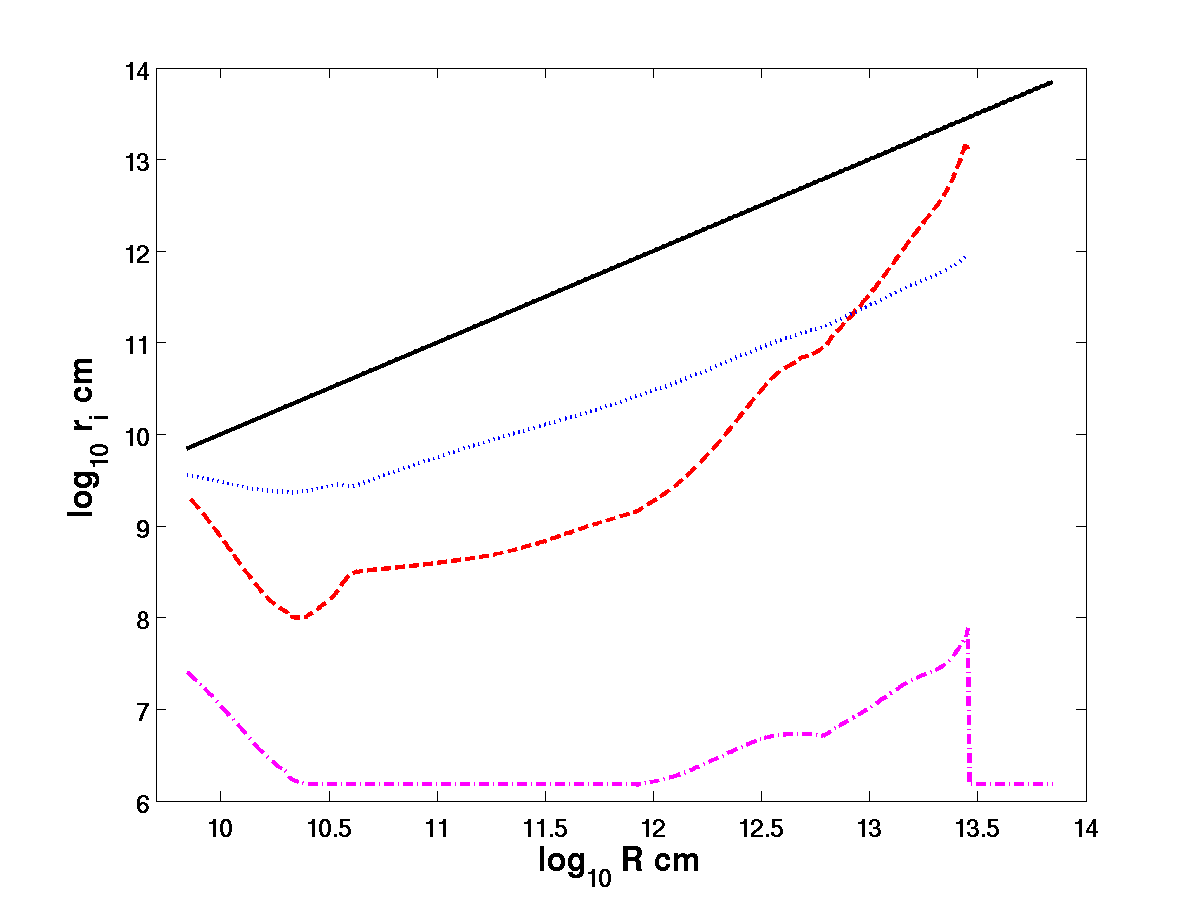}
\includegraphics[width=84mm,angle=-0]{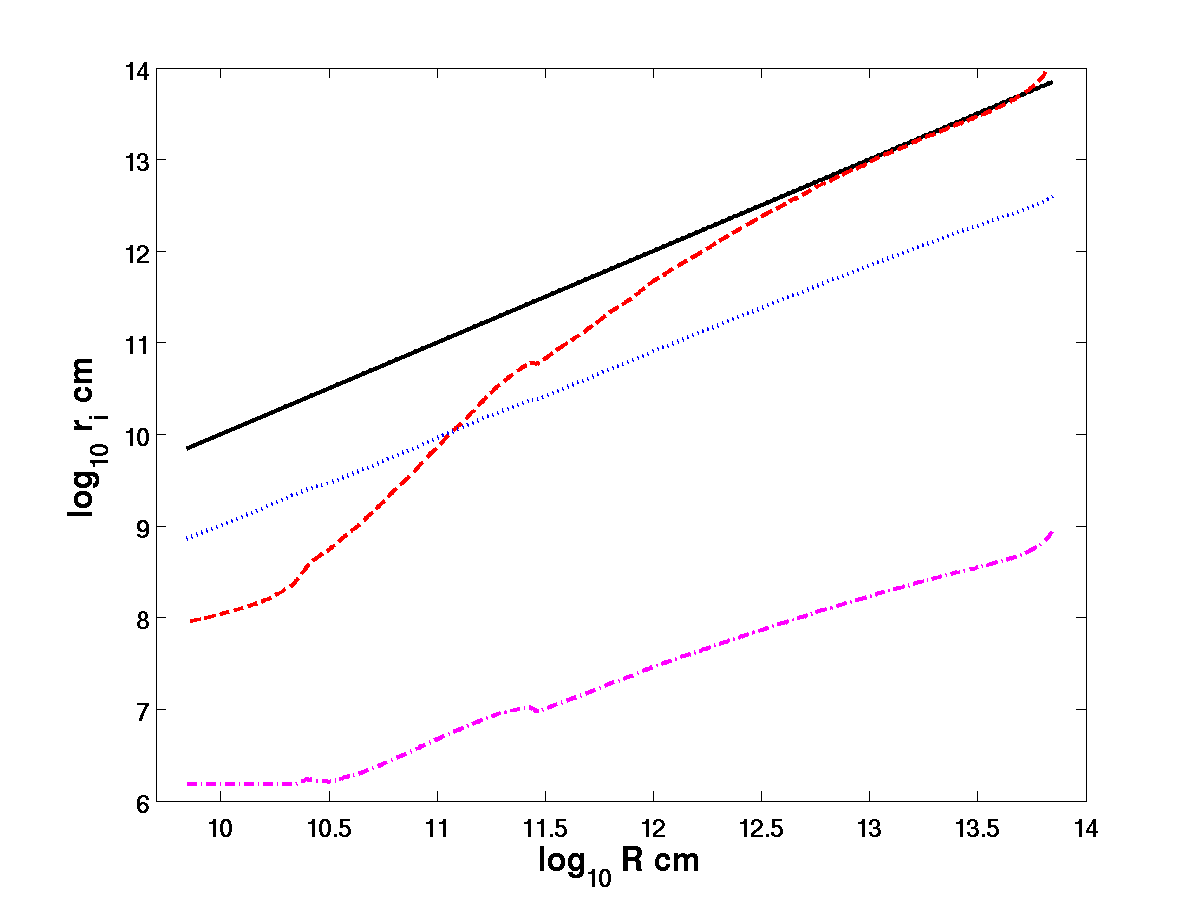}
\includegraphics[width=84mm,angle=-0]{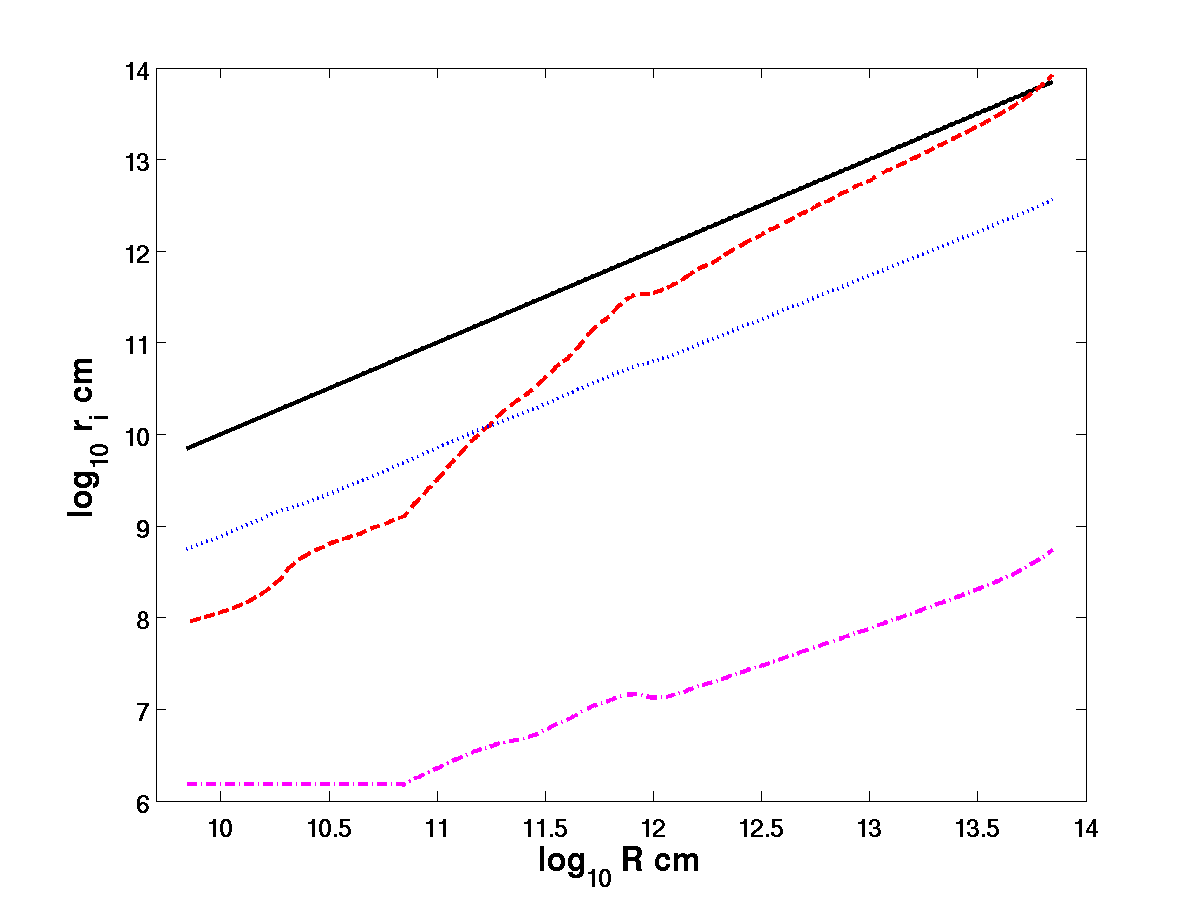}
\caption{The characteristic radii of the in-spiral problem as functions 
of the position inside the RSG primary for the same models as in Figure.\ref{dmdtf} 
and $\eta=0.1$.  
The orbital radius is shown by the solid line, the sonic point, $R\sub{acc}$, by 
the dotted line, the accretion shock radius by the dashed line,  and the
circularisation radius, $R\sub{c}$ by the dash-doted line. 
}
\label{rcyr}
\end{figure*}

Figure~\ref{dmdtf} shows the Bondi-Hoyle and the critical accretion rates for these 
six models and the two extreme values of the parameter $\eta$, representing 
the cases of both very efficient ($\eta=1$) and rather inefficient ($\eta=0.1$) 
accretion of angular momentum (see Eq.\ref{ja}). One can see that 
the condition of radiation trapping is not very restrictive. 
For $\eta=1$ it is satisfied for $R<3\by10^{11}\,$cm in models RSG20 and RSG25, 
for $R<3\by10^{12}\,$cm in models RSG20He and RSG25He, and much earlier 
for $\eta=0.1$.  

On the contrary, the condition $\dot{M}\sub{BH}>\dot{M}\sub{cn2}$ can be quite 
restrictive. In fact, it is never satisfied when $\eta=1$. In this case, the 
circularisation radius is always too far away from the NS and the neutrino 
cooling is not efficient enough.  For $\eta=0.1$ the results seem more promising
as this condition is satisfied long way before NS merges with the core. 
For RSGs in the middle of the He-burning phase 
this occurs almost the the same point where the radiation becomes trapped, and  
a bit later for the presupernova RSGs. Thus, for $\eta=0.1$ NS begins to hyper-accrete 
already inside the envelope of the RSG primary.
However, for such a small value of $\eta$ the accretion disk becomes very 
compact and may even disappear, in which case the recycling efficiency 
is reduced.        
 
Figure~\ref{rcyr} shows all the characteristic radii of the in-spiral problem 
for $\eta=0.1$. One can see that for RSG20 and RSG25 models, the disk accretion 
domain has  $a > 3\by10^{10}\,$cm.  
For the model RSG20He this domain includes the whole stellar envelope and 
for RSG25He it splits into two zones. What is most important is that in 
all models the accretion begins to proceed with the Bondi-Hoyle rate while 
still in the disk regime.

Once the accretion disk is formed the specific angular momentum of the gas  
settling on the NS surface is that of the last Keplerian orbit. In this case 
the star will reach the rotational rate $\Omega$ after accumulating  
\begin{equation} \Delta M \simeq \frac{\Omega I}{j\sub{K}} 
\simeq 0.18\, M\sub{0} R\sub{NS,6}^2 P\sub{-3}\, M_{\odot}
\label{dm}
\end{equation}
where $I\simeq(2/5)MR\sub{NS}^2$ is the final moment of inertia of NS 
and $j\sub{K}$ is the Keplerian angular momentum at the NS radius.
Obviously, the same mass is required to recycle the observed millisecond pulsars 
in low mass binaries and this allows us to conclude that the NS may well reach 
the gravitationally unstable rotation rate before collapsing into a black hole. 

\begin{figure*}
\includegraphics[width=58mm,angle=-0]{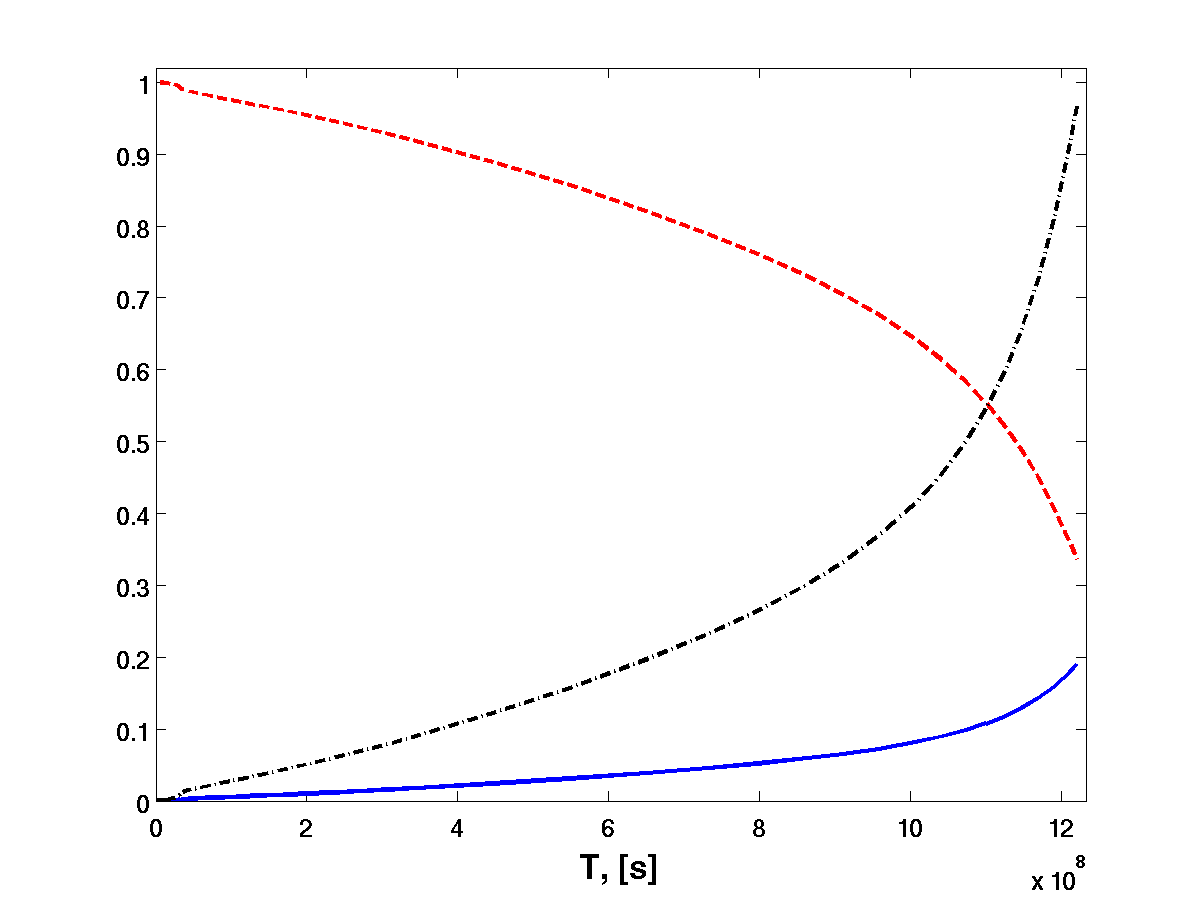}
\includegraphics[width=58mm,angle=-0]{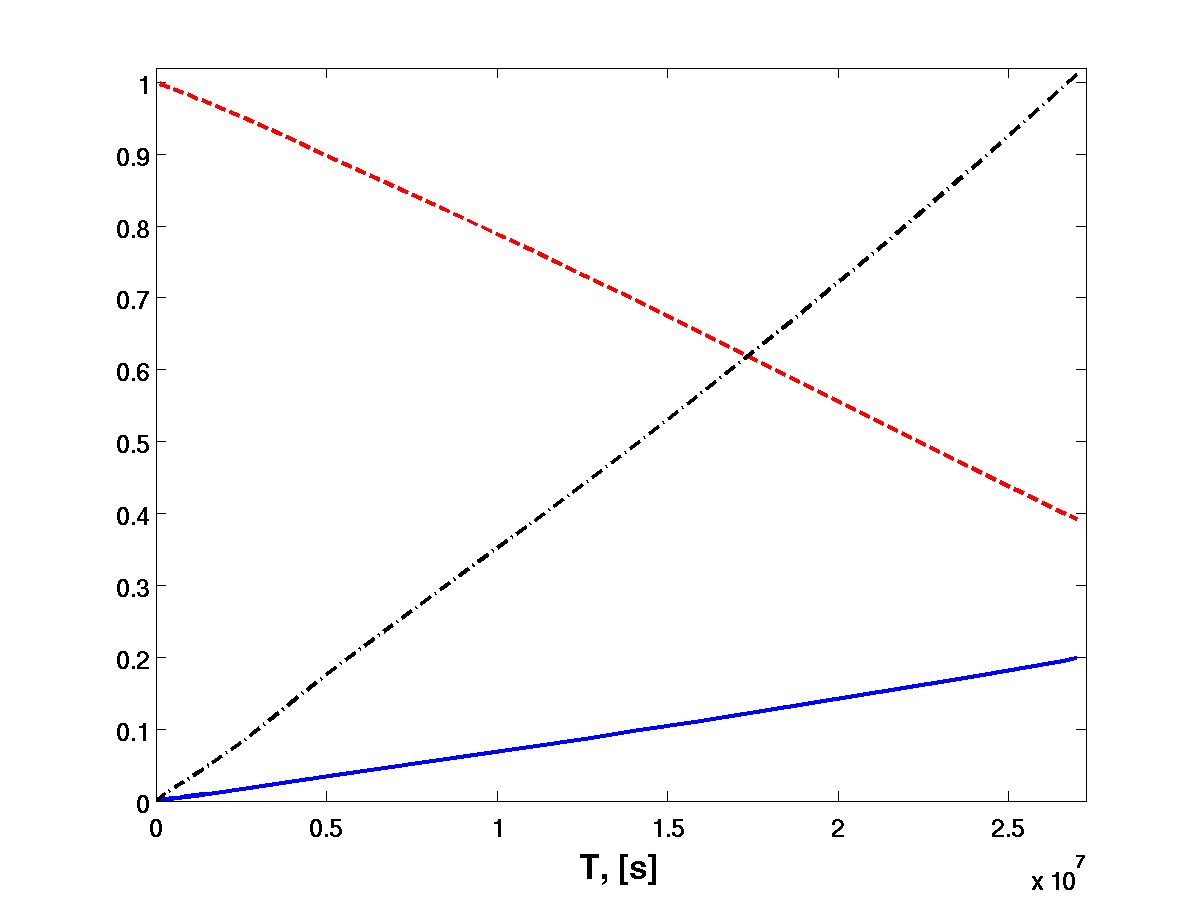}
\includegraphics[width=58mm,angle=-0]{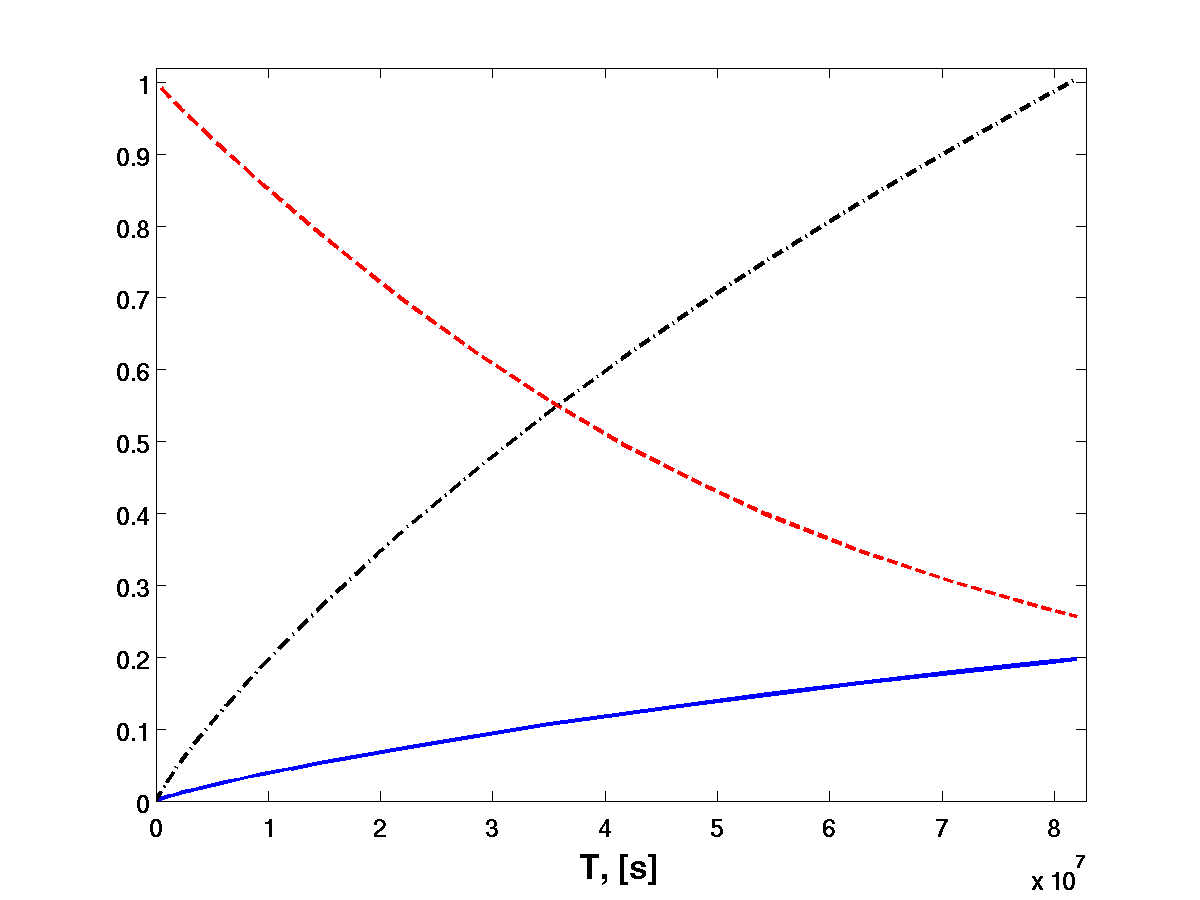}
\caption{Evolution of NS in the case when the Bondi-Hoyle regime begins at 
the surface of the RSG primary for the models RSG25He (left panel), 
 RSG20 (middle panel), and  RSG25 (right panel).  
The lines show the accreted mass normalized to $M_{\odot}$ (solid line), the 
angular momentum of NS normalized to critical value $j_c=I\Omega_c$, where 
$\Omega_c=760\,$Hz (dot-dashed line), and the orbital radius normalized to its 
initial value (dashed line). 
}
\label{f_amj}
\end{figure*}

\begin{figure*}
\includegraphics[width=58mm,angle=-0]{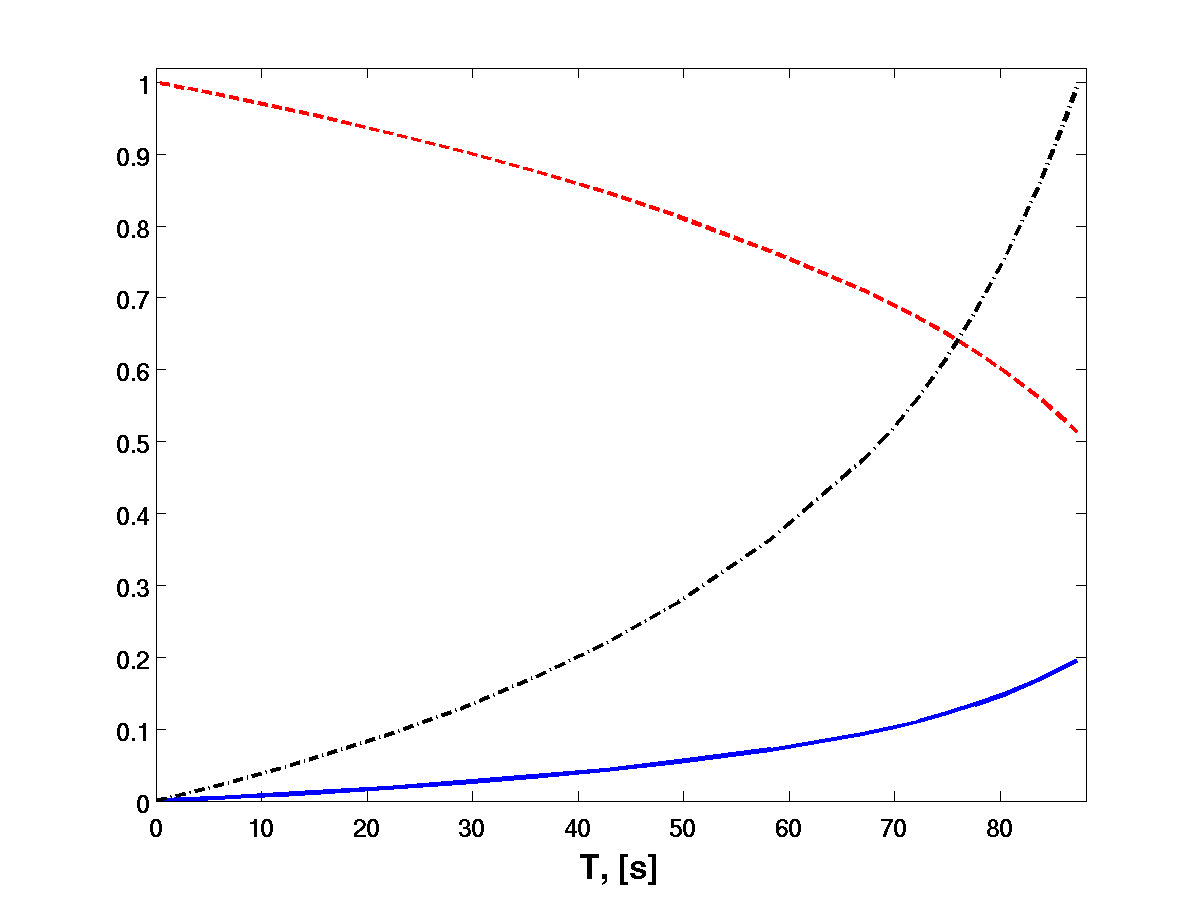}
\includegraphics[width=58mm,angle=-0]{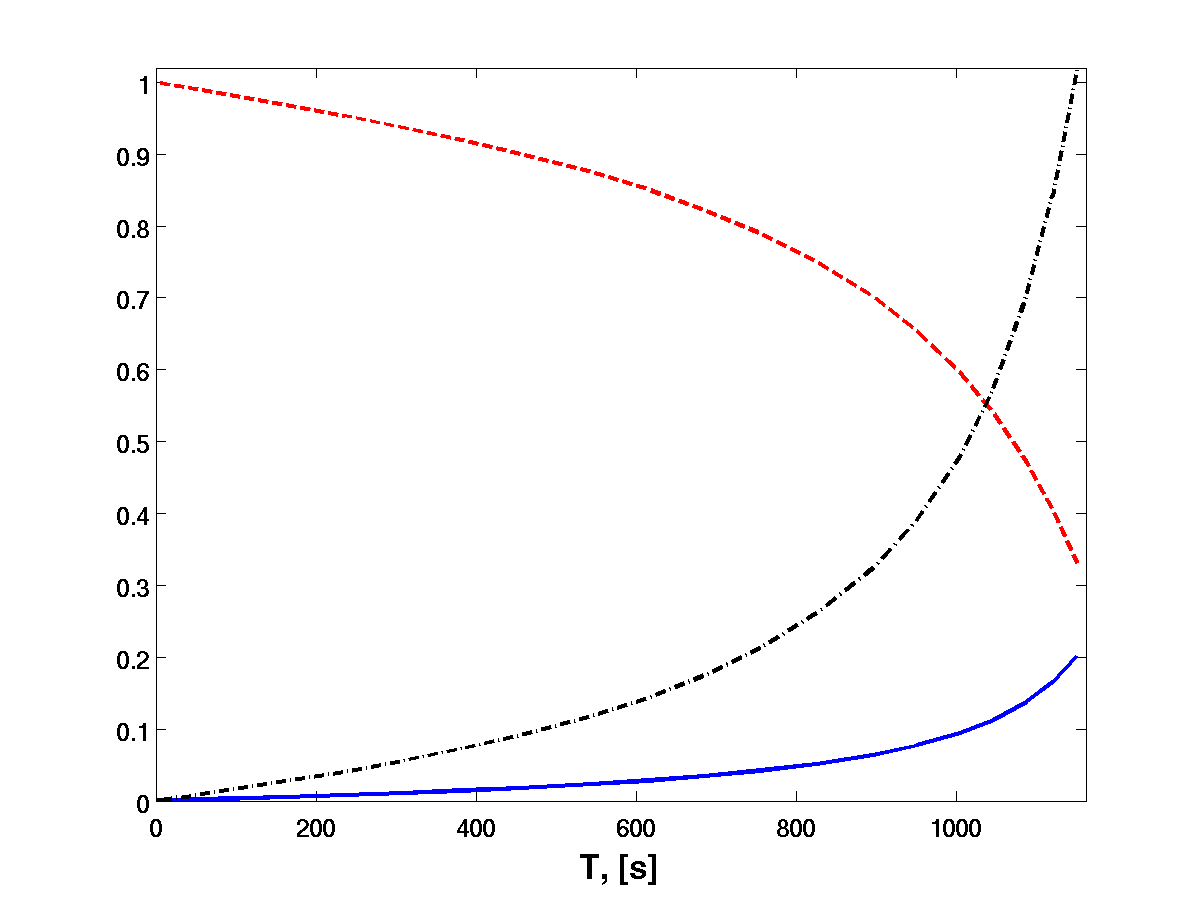}
\includegraphics[width=58mm,angle=-0]{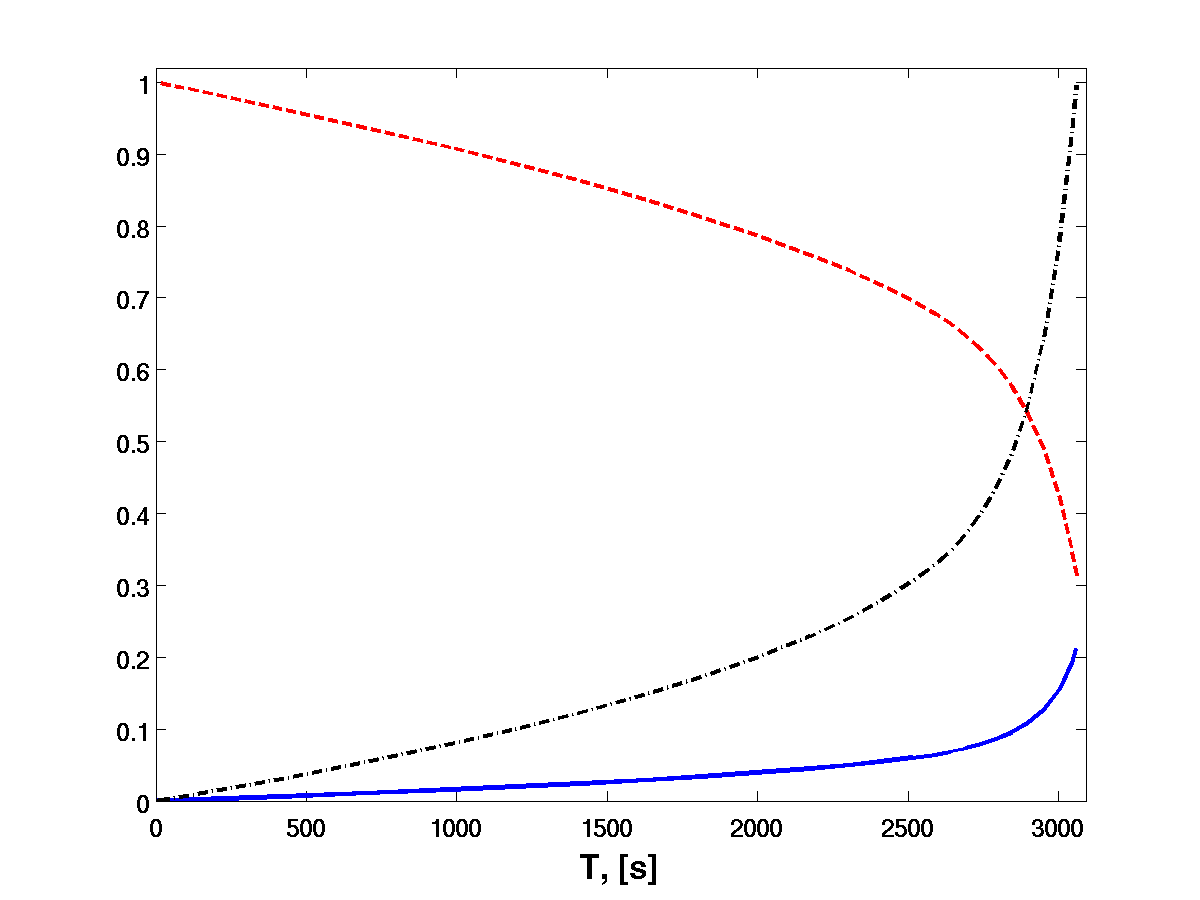}
\caption{The same as in Figure~\ref{f_amj} for the case when the Bondi-Hoyle 
regime starts at $a=3\by 10^{10}\,$cm. }
\label{f_amjin}
\end{figure*}

When the accretion proceeds at the Bondi-Hoyle rate, the NS mass and orbit 
are related approximately as  
\begin{equation} 
\frac{a\sub{end}}{a\sub{init}}=\left(\frac{M\sub{init}}{M\sub{end}}\right)^\sigma,  
\end{equation} 
where $\sigma=-M\dot{a}/\dot{M}a$. \citet{ch93} found that for 10$M_\odot$ RSG 
$\sigma$ varies slowly between 5 and 7. According to these results, NS increases 
its mass by $0.2M_\odot$ after its orbital radius decreases by only a 
factor of few.  This shows that the ``window of opportunity'', 
where the disk accretion can proceed at the Bondi-Hoyle rate, does not 
have to be particularly wide.   
In order to obtain more accurate estimate, one could integrate 
the evolution equation  
\begin{equation} 
\frac{\dot{a}}{a}=-\frac{4\pi G^2M \rho}{v^3}
\left[\frac{M}{M_*(1+{\cal M}^{-2})^{3/2}}+\zeta C_D\right]
\left(\zeta+3\frac{\rho}{\overline{\rho}}\right)^{-1},
\label{dadt}
\end{equation} 
where $\overline{\rho}$ is the average density of the primary inside $r=a$,
$C_D\approx 6.5$ is the drag coefficient\cite{smts85}, and $\zeta=1+M_N/M_*$ \citep{ch93}.
Figures~\ref{f_amj} and \ref{f_amjin} show the results of such integration for the models 
RSG20, RSG25, and RSG25He, assuming that the hyper-accretion regime begins near the stellar 
surface and at $a=3\by10^{10}\,$cm respectively. One can see that in all these cases 
the gravitationally unstable rotation rate is reached when $a$ decreases by less  
than a factor of five, in good agreement with the original estimate.

For $\eta=1$, the mass accretion rate is low and the NS cannot spin-up 
significantly inside the CE. What occurs in this case depends on the details of 
CE evolution. The CE can either be ejected leaving behind a close NS-WR binary or 
survive, leading to a merger of the NS and the RSG core \citep{Taams00}.  
The configuration after the merger is similar to that of the Thorne-Zydkow object 
\citep{tz77,bbl01}. The common 
view is that such objects can not exist because the effective neutrino 
cooling leads to hyper-accretion and prompt collapse of the NS into a BH.  
This conclusion is based on the spherically symmetric model, 
where the circularisation radius is zero and the flow stagnation surface coincides 
with the surface of the NS. However, during the inspiral the orbital angular momentum 
of the NS is converted into the spin of the primary.  Thus, one would 
expect the core to be tidally disrupted and form a massive accretion disk around the NS. 
As the result, the cooling rate will be lower compared to that found in the spherically 
symmetric case.  It is not obvious for how long this configuration may exist. 
However, this again opens the possibility of recycling of the NS up to the gravitationally 
unstable rotation rate and magnetically driven stellar explosion.

In the case of core-collapse, even the magnetic field of magnetar strength 
cannot prevent accretion and fails to drive stellar explosion \citep{KB07}. 
The magnetosphere becomes locked under the pile of accreting matter, 
and a separated mechanism, e.g. the standard neutrino-driven explosion, 
is required to release it.   
However, the density around the Fe core of presupernovae is very high, leading to 
very high accretion rates. In our case the accretion rates are expected to be 
significantly lower. In the next section we explore if they are low enough to allow 
purely magnetically-driven explosions.

\section{Magnetic explosion}
\label{magnetic explosion}

\subsection{Magnetic field generation}

The main route to formation of ``magnetars'' is believed to be the core collapse 
of rapidly rotating stars. During the collapse, the proto-neutron star 
naturally develops strong differential rotation, with the angular velocity decreasing 
outwards.  This allows generation of super-strong magnetic field via the 
$\alpha$-$\Omega$-dynamo \citep{DT92,TD93} or magneto-rotational instability 
\citep{bdl07}.

In our case, the conditions are rather different. The neutron star is already fully-developed, 
with a solid crust, when it enters the envelope of its companion.    
The typical NS crust has the thickness $h\approx10^5$~cm and 
the density at the bottom $\rho\sim 10^{14} \mbox{ g cm}^{-3}$ \citep[see e.g.][ ]{gyp01,lp07}.
The neutron star may have already been spinned up via accretion 
from the stellar wind, but but its rotation rate may still be well below the critical for 
development of the gravitational wave instability. In any case,    
the accreted gas of companion's envelope forms a very rapidly rotating layer above the crust.
As long as the crust exists, this layer and the NS core are basically decoupled, and interact mainly 
gravitationally. A strong tangential discontinuity exists at the outer boundary of the crust 
\citep{is99,is10}.  

As the mass of the outer layer increases, the crust pressure goes up. Because of the degeneracy of 
the crust matter, its pressure depends mainly on its density, and only weakly on temperature.
When the density reaches the critical value  $\rho_{cr} \sim 10^{14} \mbox{ g cm}^{-3} $,  
the crust begins to melt \cite{brown00}, and the barrier separating the NS core and its outer 
rotating layer disappears. If we ignore the centrifugal force than this occurs when the 
NS accumulates mass comparable to the crust mass, $\sim 0.05 M_{\odot}$.  
The centrifugal force significantly reduces the effective gravitation acceleration 
and the above estimate for the layer mass is only a lower limit. Assuming  
$v_{\phi}\approx v_K\sim 10^{10} \mbox{cm s}^{-1}$ the rotational energy of the layer at the 
time of melting may well reach $\sim 10^{52}$~ergs. Once the crust has melted, the core and the 
layer can begin to interact hydromagnetically, with up to $E \sim 10^{52}$~ergs of energy in 
differential rotation to be utilised. 

One possibility is viscous heating and neutrino emission.  
We can estimate the neutrino luminosity as $L_{\nu}\approx E/t_{d}$, where $t_{d}$ is the 
neutrino diffusion time. Following the analysis of \citet{lp07} and \citet{b08}, 
this time can be estimated as
\begin{equation} 
t\sub{d}\sim h\tau_{\nu}/c,
\label{taud}
\end{equation} 
where $\tau_{\nu}=h\rho \Sigma_{\nu}/m_p$ is the optical depth of the hot envelope for 
the neutrino emission.
Assuming equilibrium with the electron fraction $Y_e=0.05$,
the mean neutrino cross-section is about 
$\Sigma_{\nu}\approx\sigma_{\nu}\epsilon_{\nu}kT_e/m_e^2c^4$, 
where $\sigma_\nu = 1.7\times 10^{-44} \mbox{ cm}^2$ \cite{bah64} and   
$\epsilon_{\nu} \approx 3.15 kT_e$ is mean energy of neutrino \citep{tbm01}. 
The temperature can be estimated as 
$$
T_e\approx (E/4\pi a_r h R\sub{NS}^2)^{1/4}\approx 
10^{12} E_{52}^{1/4}h_5^{-1/4} R\sub{NS,6}^{-1/2}\,\mbox{K},
$$
where $a_r=7.564\times10^{-15} \mbox{ erg cm}^{-3} \mbox{K}^{-4}$ is the radiation 
constant. The corresponding diffusion time is quite short, 
$$
t\sub{d}\sim 0.1\, h_5^{3/2} \rho_{14} E_{52}^{1/2}/R\sub{NS,6} \,\mbox{s}.
$$ 

Another process is development of the Kelvin-Helmholtz instability, production of turbulence and 
amplification of magnetic field. For the Kelvin-Helmholtz instability to develop the Richardson
number $J$ has to be below than 1/4 \citep{cha61}. In our case   
\begin{equation} 
J\approx \frac{gh}{v_{\phi}^2} \approx 0.2 R\sub{NS,6} h_{5} 
\label{Ri}
\end{equation} 
where $g\approx GM_{NS}/R_{NS}^2$ is the gravitational acceleration and we use 
$M_{NS}=1.5 M_{\odot}$. 
Thus, the instability condition is marginally satisfied. Once the turbulence has developed the 
magnetic field amplification is expected to proceed as discussed in \cite{DT92}.     
The e-folding time is given by the eddy turn-over time, which one would expect to be below the 
rotation period of the outer layer. Thus, strong magnetic field can be generated on the time 
below the viscous and the neutrino diffusion time scale\footnote{The viscous dissipation 
timescale is very long  
$t_{vis}\approx10^8 R_{NS,6}^{23/4}T^2_9$ s \citep{shap00}.}
This conclusion is supported by the recent numerical simulations
of similar problems, which show that the time as short as 0.03 s can be sufficient to 
generate dynamically strong magnetic field \citep{ocma09,rgb11}. The strong toroidal magnetic field
can suppers the instability in the case if $\rho v^2<B^2/4\pi$ \citep{cha61} or if $B>3\times 10^{17}$~G.
This is likely to determine the saturation strength of magnetic field. Such a strong magnetic 
field is also buoyant \citep{s99} and will emerge from under the NS surface into the 
surrounding accretion flow. 

Another possibility of generating super-strong magnetic in recycled NS was proposed by 
\citet{s99} with application to X-ray binaries. 
In this model, the NS spins up to the critical rotation rate where it becomes 
unstable to exciting r-modes and gravitational radiation \citep{l99,s99}. This may lead to 
effective braking of the outer layers of NS and developing of strong differential rotation.  
Recent studies, however, suggest that non-linear coupling with other modes leads to the 
saturation of the r-mode at a much lower amplitude \citep{afm03,btw05,btw07}, though the mass 
accretion rates assumed in these studies are much lower than those encountered in common
envelopes.   

\subsection{Criteria for explosion}

Depending on the mass accretion rate, this may just modify the 
properties of the accretion flow near the star, the ``accretor'' regime, 
or drive an outflow \citep[e.g.][]{is75}. In the latter case, two different 
regimes are normally discussed in the literature, the ``propeller'' regime and 
the ``ejector'' regime. 
In the ejector regime, the magnetosphere expands beyond the light cylinder 
and develops a pulsar wind. In the propeller regime, the magnetosphere 
remains confined within the light cylinder. If the magnetic axis is not 
aligned with the rotational axis it may act in a similar fashion to an aircraft 
propeller, driving shock waves into the surrounding gas, heating it and spinning it 
up. If the energy supplied by the propeller is efficiently radiated away, this may 
lead to a quasi-static configuration \citep[e.g.][]{mrf91}. 
If not, the most likely outcome is an outflow, followed by expansion of the 
magnetosphere, and transition to the ejector regime.

This classification is applied both in the case of quasi-spherical accretion
and the case of disk accretion. In the disk case, the external gas above and  
below the disk is usually not considered as important dynamical element. 
However in our case this is not true because the accretion disk is rather 
compact and feeds from a more or less spherical component of the accretion flow.
If the expanding  magnetosphere removes this component,
the disk will no longer be supplied  with mass and quickly 
disappear, no matter what its regime is.              
For this reason, we will not consider the complicated 
interaction between the magnetosphere and the disk\footnote{An interested reader may 
find out more about the disk-magnetosphere interaction from \citet{ukrl06} and references 
therein.}, but will focus on the interaction with       
the low angular momentum flow of the polar regions which accretes directly onto 
the NS avoiding the disk. In order to simplify the problem, we will  
treat this flow as spherically symmetric, keeping in mind that its mass 
accretion rate is only a fraction of the Bondi-Hoyle rate.

The extent of the magnetosphere is often estimated by the balance between the 
magnetic pressure and the ram pressure of accreting flow.
This is fine when the accretion shock is close to the NS. If not, the thermal 
pressure at the magnetospheric radius can be significantly higher compared both to 
the actual ram pressure at the shock and to the ram pressure that would be 
found at the magnetospheric radius if the free-fall flow could continue down to 
this radius. In this case, the initial radius of emerging magnetosphere is 
rather determined by the balance of the magnetic pressure and the thermal pressure 
of the quasi-hydrostatic ``settling'' flow downstream of the accretion shock \citep{mrf91}.    

Assuming that the accretion flow in the polar region is approximately 
spherical, we can describe it using the analytical model developed by \citet{hch91}. 
Repeating their calculations with the more accurate cooling function, adopted 
later in \citet{ch96}, 
\begin{equation}
    \dot{\epsilon}\sub{n} = 10^{25}\fracb{kT}{\mbox{MeV}}^9 
       \frac{\mbox{erg}}{\mbox{cm}^3 \mbox{s}} 
\label{cf}
\end{equation} 
\citep{dic72,bw94} and using the adiabatic index $\gamma=4/3$, 
we find the accretion shock radius 
\begin{equation}
R\sub{sh} \simeq 8.2\by10^{8} 
          f_1
          R\sub{NS,6}^{40/27} 
          M\sub{0}^{-1/27}
          \dot{M}\sub{0}^{-10/27} \mbox{cm}, 
\label{r-sh1}
\end{equation}
where 
\begin{equation}
   f_1 = \left[ \frac{R\sub{NS}}{R\sub{g}} 
        \left(\left(1-\frac{2R\sub{g}}{R\sub{NS}}\right)^{-1/2}-1\right)
         \right]^{-64/27},
\end{equation}
and $R\sub{g}=GM/c^2$ is the gravitational radius of the NS (In the calculations we 
assume the Schwarzschild space-time.). For reasonable masses and radii of NSs,  
$0.27<f_1<0.62$. In particular $f_1=0.54$ for $M=M_\odot$ and $R\sub{NS}=10\,$km. 
Downstream of the shock we have an adiabatic subsonic flow with the mass density  
\begin{equation} 
\rho = \rho\sub{sh} \fracb{R}{R\sub{sh}}^{-3},
\label{rho-hs}
\end{equation}
where 
\begin{equation}
  \rho\sub{sh}=\frac{7}{4\pi} \frac{\dot{M}}{(GM)^{1/2} R\sub{sh}^{3/2}}
\end{equation}
is the mass density at the shock, the pressure 
\begin{equation}
p = \frac{6}{7} p\sub{ram}(R\sub{sh}) \fracb{R}{R\sub{sh}}^{-4},
\label{p-hs}
\end{equation}
where
\begin{equation}
  p\sub{ram}(R\sub{sh})= 
       \frac{\dot{M} \sqrt{GM}}{4\pi R\sub{sh}^{5/2}}
\end{equation}
is the ram pressure of the free falling flow at the shock, the 
temperature 
\begin{equation}
T = \left(\frac{12}{11a} p\right)^{1/4},
\label{T-hs}
\end{equation}
where $a$ is the radiation constant, and the radial component of 
4-velocity 
\begin{equation}
u = -\frac{\dot{M}}{4\pi\rho R^2}.
\label{u-hs}
\end{equation}

Assuming that the magnetosphere expands until its 
magnetic pressure equals to the gas pressure\footnote{
As the result of its oscillations the NS can become quite hot and fill the 
magnetosphere with substantial amount of plasma. If this case the centrifugal 
force will need to be included in the force balance, yielding even higher value for 
$R\sub{m}$.} 
we find the radius of dipolar magnetosphere with the surface strength 
$B\sub{NS}$ to be 
\begin{equation}
R\sub{m} = 5\by10^6 
      f_1^{-3/4} 
      R\sub{NS,6}^{17/9} 
      M\sub{0}^{-2/9} 
      \dot{M}\sub{0}^{-2/9} 
      B\sub{NS,15} 
      \,\mbox{cm}. 
\label{rm-d}
\end{equation}  
For this to be above the NS radius the mass accretion rate should 
not exceed 
\begin{equation}
   \dot{M}\sub{cr} \simeq 1.3\by10^3 
        f_1^{-27/8}
        B\sub{NS,15}^{9/2}
        R\sub{NS,6}^{4} 
        M\sub{0}^{-1}
         \frac{M_\odot}{\mbox{yr}}.
\label{m-cr1a}
\end{equation}

The traditional criterion for the propeller regime is that 
the magnetospheric radius is below the light cylinder radius 
\begin{equation}
R\sub{LC}=cP/2\pi \simeq 4.8\by10^6 P\sub{-3} \, \mbox{cm}
\label{R_LC}
\end{equation}
and above the co-rotation radius 
\begin{equation}
R\sub{cor}=\fracb{GM}{\Omega^2}^{1/3} \simeq 
         1.5\by10^6  P\sub{-3}^{2/3}
                       M\sub{0}^{1/3}
                       R\sub{NS,6}\,\mbox{cm}\, ,
\label{R_c}
\end{equation}
at which the magnetosphere rotates with the local Keplerian velocity. 

The ejector regime is realised when $R\sub{m}>R\sub{c}$. 
In terms of the mass accretion rate the propeller 
regime criterion can be written as 
$$
       \dot{M}\sub{ej}<\dot{M}<\dot{M}\sub{pr},
$$
and the ejector regime criterion as 
$$
       \dot{M}<\dot{M}\sub{ej},
$$  
where
\begin{equation}
\dot{M}\sub{pr} \simeq 2\by10^2
        f_1^{-27/8}
        M\sub{0}^{-5/2}
        R\sub{NS,6}^{17/2} 
        B\sub{NS,15}^{9/2}
        P\sub{-3}^{-3}
         \,\frac{M_\odot}{\mbox{yr}}
\label{m-crit2da}
\end{equation}
and 
\begin{equation}
\dot{M}\sub{ej} \simeq 1.1
        f_1^{-27/8}
        M\sub{0}^{-1}
        R\sub{NS,6}^{17/2} 
        B\sub{NS,15}^{9/2}
        P\sub{-3}^{-9/2}
         \,\frac{M_\odot}{\mbox{yr}}. 
\label{m-crit3da}
\end{equation}
The latter is much stronger, and even if only a fraction of the 
total mass accretion rate is shared by the quasi-spherical component, 
it could be satisfied only during the early phase of the inspiral 
(see Figure~\ref{dmdtf}).  More likely, the emerged magnetosphere 
will operate in the propeller regime. 

The sound speed in the \citet{hch91} solution depends only of the NS mass 
and the distance from the NS, 
\begin{equation}
   c\sub{s} \simeq 6.2\by10^9 M\sub{0}^{1/2} R\sub{6}^{-1/2}
                  \, \frac{\mbox{cm}}{\mbox{s}}. 
\end{equation}
The linear rotational velocity of the magnetosphere,  
\begin{equation}
  v\sub{rot} \simeq 4.8 \by10^9 P\sub{-3}^{-1} R\sub{6} 
                             \, \frac{\mbox{cm}}{\mbox{s}}\, ,
\end{equation}
is generally higher and thus we have the so-called supersonic propeller. 
The power generated by such a propeller is roughly 
\begin{displaymath}
 L\sub{pr} \simeq \frac{4\pi}{3} p\sub{m} R\sub{m}^2 
                  \frac{(R\sub{m}\Omega)^2}{c\sub{s}} 
\end{displaymath}
\begin{equation}
   \qquad  \simeq  1.3\by10^{50} f_1^{9/8}
        B\sub{NS,15}^{1/2}
        P\sub{-3}^{-2}
        \dot{M}\sub{0}^{1/3}
        M\sub{0}^{1/6}
         \,\frac{\mbox{erg}}{\mbox{s}}
\label{L-pr}
\end{equation}
\citep{mrf91}, exceeding by several orders of magnitude the accretion power 
\begin{equation}
  L\sub{acc} =\frac{GM\dot{M}}{R\sub{NS}} 
        \simeq  8\by10^{45}
        M\sub{0}
        R\sub{NS,6}^{-1}
        \dot{M}\sub{0}
         \,\frac{\mbox{erg}}{\mbox{s}}.
\label{L-acc}
\end{equation}
This suggests that the magnetosphere easily blows away the accreting matter and  
expands beyond the light cylinder, thus creating the conditions for 
proper pulsar wind from the NS. 

One may argue that the accretion flow could re-adjust to the new conditions, 
with the accretion shock moving further out, the temperature at the 
magnetospheric radius rising, and the enhanced neutrino cooling compensating 
for the propeller heating, $L\sub{n} = L\sub{pr}$ (A similar problem has been 
analysed by \citet{mrf91}.). Given the strong dependence 
of the neutrino emissivity on temperature, the cooling is expected to be 
significant only in the close vicinity of the magnetosphere. In the non-magnetic 
case, most of the cooling occurs in the volume comparable to the NS volume 
\citep{hch91}. This suggests that in our case the cooling volume will be approximately 
the same as the volume of the magnetosphere. Then the energy balance can be written as 
\begin{equation} 
    \frac{4\pi}{3} R\sub{m}^3\dot{\epsilon}\sub{n}(T) = L\sub{pr},  
\end{equation}         
where $\dot{\epsilon}\sub{n}(T)$ is given by Eq.\ref{cf}. Solving this equation for 
$R\sub{m}$, simultaneously with the pressure balance 
\begin{equation}
     p(R\sub{m})= \frac{1}{8\pi} B_0^2 \fracb{R\sub{m}}{R\sub{NS}}^{-6},
\end{equation}
where $p(R)$ is given by Eq.\ref{p-hs}, we find that  
\begin{equation}
     \frac{R\sub{m}}{R\sub{NS}} \simeq 0.6 
        B\sub{NS,15}^{5/18} 
        R\sub{NS,6}^{-1/6}
        M\sub{0}^{1/18}
        P\sub{-3}^{2/9}, 
\end{equation}
independently on the mass accretion rate. Since in this solution 
$R\sub{m}<R\sub{NS}$, we conclude that the accretion flow cannot re-adjust
itself and will be terminated by the emerged magnetosphere.   

In the case of cold neutron stars, the pulsar wind power  
\begin{equation}
  L\sub{w} \simeq \frac{\mu^2\Omega^4}{c^3}(1+\sin^2\theta\sub{m}) 
        \simeq  6\by10^{49} 
         B\sub{NS,15}^2 
         R\sub{NS,6}^{6}
         P\sub{-3}^{-4}
         \,\frac{\mbox{erg}}{\mbox{s}},
\label{L-w}
\end{equation}
where $\mu = B_0 R\sub{NS}^2$ is the star magnetic moment and $\theta\sub{m}$ is the 
angle between the magnetic moment and the rotational axis of the NS \citep{spt06}. 
For a hot NS the mass loading of the wind can be 
substantial, leading to even higher luminosity \citep{bucc06,mtq07}. The energy 
released in the magnetically driven explosion is the rotational energy of the NS 
\begin{equation}
  E\sub{rot} = \frac{1}{2} I \Omega^2  \simeq
        1.6\by10^{52}
        M\sub{0}
        R\sub{NS,6}^{2}
        P\sub{-3}^{2}
        \,\mbox{erg}\, ,
\label{E-r}
\end{equation}    
where $I\simeq (2/5)M\sub{NS}R\sub{NS}^2$ is the NS moment of inertia.

Inside the RSG core the accretion rate is expected to be much higher.
For example, in the RSG20He model the core density
$\rho\simeq 8.6\by10^2\mbox{g}/\mbox{cm}^{3}$ and the pressure 
$p\simeq1.3\by10^{19}\mbox{g}\,\mbox{cm}^2/\mbox{s}^2$,
yielding the Bondi accretion rate
\begin{equation}
   \dot{M}\sub{B} \simeq \pi \frac{(GM)^2 \rho}{c_s^3}
    \simeq 4.3\by10^5 \, \frac{M_\odot}{ \mbox{yr}}.
\label{Bondi}
\end{equation}
For such a high accretion rate, the propeller regime requires the magnetic field
to be well in excess of $10^{15}$G, similar to what is required in the magnetar 
models of GRBs. Whether such a strong dipolar magnetic field 
can be generated is not clear at present, as the observations of known magnetars 
suggest lower values. However, this could be a result of the short decay time 
expected for such a strong field \citep{hk98}.

\section{Supernova properties}
\label{Discussion}

\subsection{The plateau.}

Since the NS has to reach the maximum allowed rotation rate before 
it explodes, the model predicts the explosion energy around $10^{52}$erg, 
putting it on the level of hypernovae. The corresponding ejecta speed will 
be very high 
\begin{equation}
  v\sub{ej}=10^9 M\sub{ej,1}^{-1/2}\, \frac{\mbox{cm}}{\mbox{s}}\, , 
\end{equation}
where $M\sub{ej,1}$ is the ejected mass.

Since the progenitor is a RSG the supernova spectra will show hydrogen 
lines and would be classified as type-II. 
As the NS spirals through the weakly bound outer envelope of its RSG 
companion, it drives an outflow with the speed about the local 
escape speed and the mass loss rate about $M_\odot/$yr \citep{Taams00}. 
The total in-spiral time is about few initial orbital periods. 
For a RSG with mass $20\,M_\odot$ and radius $R=10^{14}$cm this 
yields $v\sub{esc}=5\by10^6 {\rm cm}/{\rm s}$ and $t\sub{ins}\sim\mbox{few}\times10^8$s. 
Thus, we expect a hydrogen rich circumstellar shell of few solar masses to 
exist at a distance of $\sim 10^{15}$cm at the time of the supernova 
explosion.  The presence of such a shell will delay the transition to adiabatic 
expansion phase and increase the supernova brightness similar to how this occurs 
type-IIn supernovae \citep{cbc04,wbh07,slf07,scl08}.  

Following \citet{kb10}, the supernova luminosity during the phase of adiabatic 
expansion can be estimated as 
\begin{equation}
   L \simeq \frac{E\sub{rot} t\sub{e}}{t\sub{d}^2},   
\end{equation}
where $t\sub{e}=R\sub{0}/v\sub{ej}$ is the expansion time scale and $R\sub{0}$ is 
the progenitor radius, and 
\begin{equation}
      t\sub{d} = \left( 
          \frac{3}{4\pi} \frac{M\sub{ej}k}{cv\sub{ej}}
      \right)^{1/2}
\end{equation}
is the radiative diffusion time scale, $k$ is the opacity (In the estimates below we use 
$k=0.4\,\mbox{cm}^2/\mbox{g}$, the value for Thomson scattering in fully ionised 
plasma.). It is assumed that 
the initial magnetar spindown time scale is smaller compared to 
$t\sub{e}$, which is certainly satisfied in our case, given the large required 
magnetic field, $10^{15}-10^{16}$G.   For the characteristic parameters 
of the model we have 
\begin{equation}
 t\sub{e} \simeq 10^5 R\sub{0,14}v\sub{ej,9}^{-1}\, \mbox{s},\quad
 t\sub{d} \simeq 8\by10^6 M\sub{ej,1}^{1/2} v\sub{ej,9}^{-1/2}\, \mbox{s}, 
\end{equation}
and 
\begin{equation}
  L \simeq 1.6\by10^{43} R\sub{0,14} M\sub{ej,1}\, 
 \frac{\mbox{erg}}{\mbox{s}}. 
\end{equation}
This high luminosity will be sustained for up to $t\sim t\sub{d}$, after 
which the trapped radiation, generated by the blast wave when it was crossing 
the star, escapes and the luminosity drops. It most normal type-II supernovae 
the luminosity after this point is sustained via radioactive decay of isotopes, 
mainly $^{56}$Ni, produced by the supernova shock in the high density 
environment around the collapsed core. However, in our model the stellar 
structure is significantly different from normal pre-supernovae. 

\subsection{$^{56}$Ni production}

If the explosion occurs after the NS is settled into the center of RSG20He  
the core density is only $\rho\sub{c}\simeq 8\by10^2 \mbox{g}\,\mbox{cm}^{-3}$ and 
it temperature $T\sub{c}\simeq 1.9\by10^8\,$K. 
where $R\sub{A} = GM/c\sub{s}^2$ is the Bondi accretion radius. 

As the magnetically driven explosion develops a shock wave propagates 
through the body of RSG. Where the post-shock temperature exceeds 
$T\sub{Ni}\simeq10^9$K, $^{56}$Ni is produced in abundance \citep{whw02}.  
In order to capture the shock dynamics, we need to know the structure of 
the accretion flow onto the NS, which has been established before the explosion. 
This flow is rather complicated because of its rotation and without computer 
simulations we can only make very rough estimates.    

its rotation and near the NS there is an accretion disk. 
First consider the polar region where the flow is more of less spherically 
symmetric. Equation~\ref{rsh} shows that the accretion shock is quite close to 
the NS, particularly when it is near the RSG core, and hence most of the 
flow can be described using the free fall model. 
The ram pressure of the accretion flow in this model is given by   
\begin{equation}
    \rho v\sub{ff}^2 = \frac{\dot{M}\sqrt{2GM}}{4\pi} R^{-5/2},
\end{equation}
where $v\sub{ff}=(2GM/R)^{1/2}$ is the free fall speed. 
Close to the NS the speed of explosion shock $v\sub{sh}$ is much less 
compared to $v\sub{ff}$ and can be found from the pressure balance equation
\begin{equation}
  \frac{2}{\gamma+1}\rho v\sub{ff}^2 = \frac{E}{4\pi R^3}. 
\end{equation}
where $R$ is the shock radius and $E=L\sub{w} t$ is the energy 
injected by the magnetar.  This yields 
\begin{equation}
    t = \frac{2\dot{M}\sqrt{2GM}}{(\gamma+1)L\sub{w}} R^{1/2}
\label{R(t)} 
\end{equation}
and thus the shock speed increases with the distance. This approximation breaks down 
at the radius
\begin{equation}
R\sub{t} = \frac{2\dot{M}GM}{(\gamma+1)L\sub{w}},  
\end{equation}
where $v\sub{sh} = v\sub{ff}$. 
The gas temperature exceeds $T\sub{Ni}$ inside the radius $R\sub{Ni}$ 
determined by the equation 
\begin{equation}
   \frac{2}{\gamma+1}\rho v\sub{ff}^2 = \frac{a}{3} T\sub{Ni}^4,
\end{equation}
and the total mass of produced $^{56}$Ni can be estimated as
\begin{equation}
  M\sub{Ni} = \dot{M} t\sub{Ni},    
\label{Ni-mass}
\end{equation}
where $t\sub{Ni}$ is the time required for the shock to reach $R\sub{Ni}$ 
according to Eq.\ref{R(t)}.  From these equations we find that 
\begin{equation}
  R\sub{t} = 10^8 \dot{M}\sub{6} L\sub{w,50}^{-1} \,\mbox{cm},
\end{equation}
\begin{equation}
  R\sub{Ni} = 7.8\by10^7 \dot{M}\sub{6}^{2/5} \,\mbox{cm}, 
\end{equation}
\begin{equation}
  M\sub{Ni} = 0.003M_\odot\, \dot{M}\sub{6}^{11/5} L\sub{w,50}^{-1}.
\end{equation}
Thus only a relatively small amount of $^{56}$Ni can be produced 
even for the highest mass accretion rates allowed in this model. 

In the equatorial plane, the accretion shock can be pushed all the way 
towards $R\sub{acc}$ (see Section \ref{In-spiral}).  Downstream of the 
shock the flow is highly subsonic with 
\begin{equation}
  \rho = \rho_0 \fracb{R}{R_0}^{-3}. 
\end{equation}
The highest density is achieved when the accretion shock is infinitely weak 
and this law is applied up to the sonic point $R\sub{acc}$. In this case, 
we can put $R_0=R\sub{acc}$ and $\rho_0=\rho_*$, the mass density of the 
RSG. The equation for shock speed is now 
\begin{equation}
  \frac{2}{\gamma+1}\rho v^2 = \frac{E}{4\pi R^3},
\end{equation}
which yields the shock radius 
\begin{equation}
    R=\frac{2}{3}\fracb{(\gamma+1)L\sub{w}}{8\pi \rho_* R\sub{acc}^3}^{1/2} 
      t^{1/2}.
\end{equation}
$R\sub{Ni}$ is now determined by the equation 
\begin{equation}
   \frac{2}{\gamma+1}\rho v^2 = \frac{aT\sub{Ni}^4}{3},
\end{equation}
where the mass of produced $^{56}$Ni is still given by Eq.\ref{Ni-mass}. 
Combining these equations we find 
\begin{equation}
  M\sub{Ni} \propto \dot{M} R\sub{acc}^{9/7} \rho_*^{3/7} L\sub{w}^{1/7}.  
\end{equation}
The highest values for $M\sub{Ni}$ are obtained inside the RSG core. 
For the RSG20He, where $R\sub{acc}\sim 5\by10^9$cm, 
$\dot{M}\sim 10^6M_\odot/\mbox{yr}$ (see Figures \ref{dmdtf} and \ref{rcyr}), 
and $\rho_*\sim 5\by10^2\mbox{g}/\mbox{cm}^3$, we find 
$M\sub{Ni}\simeq 0.02M_\odot$. Although significantly higher, this is 
still small compared to the mass deduced from the observations of 
most supernovae. 

If the density behind the shock wave is lower than $10^{6}$ g cm$^{-3}$,  
which corresponds to $\dot{M}< 10^6M_\odot/\mbox{yr}$,  
then the photon dissociation of Ni$^{56}$ becomes important as well \citep{mth10}. 
Thus, our estimation gives only an upper limit on the amount of produced Ni$^{56}$. 

Another potential source of $^{56}$Ni is the wind from the accretion 
disk itself, like in the collapsar model of GRB. Without a robust 
disk model, it is difficult to figure out how effective this mechanism is.  
For example, in the numerical simulations of \citet{mw99}, the outcome was dependent 
of the rate of introduced ``viscous'' dissipation. In any case, the much lower 
accretion rates encountered in our model suggest that the amount of $^{56}$Ni 
will be lower too. Thus, we conclude that our model predicts significant deficit of 
$^{56}$Ni.     

\subsection{The long term effects of magnetar wind}

It has been suggested recently that the long-term  
activity of magnetars born during the core-collapse can power the supernova 
light curves, mimicking the effect of radioactive decay \citep{w10,kb10}. 
In these studies, it was assumed that the rotational energy of the magnetar, 
released at the rate $\propto t^{-2}$, is converted into heat at the base of 
the supernova ejecta. One mechanism of such heating is purely hydrodynamical, 
via the forward shock wave involved in the collision of the fast magnetar wind with 
the slower supernova ejecta. In our case, however, this shock is the supernova 
shock and it is not present since the break-out. 
The other mechanism, heating by the radiation produced at the termination 
shock of the magnetar wind, remains in force and here we explore its efficiency.  

Combining Eqs.\ref{L-w} and \ref{E-r}, it is easy to obtain the well known 
law of the asymptotic spin-down of magnetic rotators   
\begin{equation} 
\Omega=\fracb{I c^3}{2\mu^2t}^{1/2} 
   \simeq 40\, 
     B\sub{NS,15}^{-1}
     t\sub{7}^{-1/2}
     \,\mbox{s}^{-1}\, 
\label{omega}
\end{equation}
and
\begin{equation}
   L\sub{w} = \frac{1}{4} \frac{I^2 c^3}{\mu^2 t^2}
     \simeq 10^{41}\,
     B\sub{NS,15}^{-2}
     t\sub{7}^{-2}
     \,\frac{\mbox{erg}}{\mbox{s}}\, 
\end{equation}
(In this Section we use $R\sub{NS}=10^6$\,cm and $M\sub{NS}=1.5M_\odot$.) 
The last equation shows that unless $B\ll 10^{15}$G the supernova luminosity 
will drop dramatically after the plateau phase. Such evolution has been 
observed in some supernovae, for example SN1994W \citep{Tsv95,scl98}.   

In order to see how exactly this energy can be used to heat the supernova 
ejecta we can appeal to the observed properties of pulsar wind nebulae. 
For example, the study of the Crab Nebula by \citet{kc84} indicates that 
immediately behind the termination shock most of the energy is in the form of 
ultra-relativistic electrons and only a fraction of it is in the magnetic 
field and there is no good reason to expect this to be different for magnetar 
winds. However, the magnetic field of a one year old magnetar wind 
nebula (MWN) is expected to be very strong, making the synchrotron cooling time 
very short.  

One can estimate the magnetic field strength via the pressure balance at 
the termination shock 
\begin{equation}
   \frac{L\sub{w}}{4\pi R\sub{w}^2 c} = \frac{B^2}{8\pi},
\label{p-m}
\end{equation}   
where $R\sub{w}$ is the radius of termination shock. Assuming that 
$R\sub{w}\simeq v\sub{ej}t$, this yields
\begin{equation} 
B \simeq   0.3\, t\sub{7}^{-2} B\sub{NS,15}^{-1} v\sub{ej,9}^{-1} \,\mbox{G}\, .
\label{bsk}
\end{equation}

The lowest energy electrons accelerated at the termination shock of the Crab 
Nebula have the Lorentz factor $\gamma\sim10^6$ \citep{kc84}. 
In the magnetic field of the MWN, they produce synchrotron photons with energy
\begin{equation}
   E\sub{$\nu$} \simeq 100\,
         B\sub{0,15}^{-1}
         v\sub{ej,9}^{-3/2}
         t\sub{7}^{-2}
         \,\mbox{keV}\, ,
\label{e-peak}
\end{equation}
and their synchrotron cooling time 
\begin{equation}
t\sub{syn} \simeq  10^4\, t\sub{7}^{4} B\sub{NS,15}^{2} v\sub{ej,9}^{2} \,\mbox{s}\, ,
\label{t-syn}
\end{equation}
is shorter compared to the dynamical time scale for few years after the explosion. 

The observations of the Crab Nebula also show that the electrons are 
accelerated up to the radiation reaction limit \citep{dJ96,kl10}. 
Assuming that the same is true in our case we expect the synchrotron spectrum 
to continue up to $E\sups{max}\sub{$\nu$,syn}\approx 100\,$MeV and the highest 
energy of the electrons to be     
\begin{equation} 
E\sub{e}\sups{max} \simeq 100\, t\sub{7} B\sub{NS,15}^{1/2} 
                   v\sub{ej,9}^{1/2} \,\mbox{TeV}\, .
\label{emax}
\end{equation}

These electrons also cool via the Inverse Compton (IC) scattering on the 
soft photons produced by the supernova shell. The ratio of the IC to the 
synchrotron energy losses $\xi=u\sub{soft}/u\sub{B}$, where $u\sub{soft}$ 
is the energy density of the soft radiation and $u\sub{B}=B^2/8\pi$ is the 
magnetic energy density. During the plateau phase  
\begin{equation}
u\sub{soft} \simeq \frac{L\sub{soft}}{4\pi R\sub{ph}^2 c} \, 
\label{u-soft}
\end{equation} 
where $R\sub{ph}$ is the radius of photosphere.  Combining this result 
with Eq.~\ref{p-m} we find  
\begin{equation}
  \xi \simeq \frac{L\sub{soft}}{L\sub{w}} \fracb{R\sub{w}}{R\sub{ph}}^2 \, . 
\end{equation}
As we have seen, $L\sub{soft}$ can be much higher than $L\sub{w}$  
but on the other hand $R\sub{w}$ can be much lower than $R\sub{ph}$.
Thus, it is hard to tell whether the IC emission can compete with the synchrotron 
one without detailed numerical simulations\footnote{
Should the  IC losses dominate, $E\sub{e}\sups{max}$ and  $E\sups{max}\sub{$\nu$,syn}$ 
are to be reduced by the factors $\xi^{1/2}$ and $\xi$ respectively.}.  After the 
plateau phase $u\sub{soft}$ drops dramatically and we expect the synchrotron losses 
to be dominant.

For photons with energy $E\sub{$\nu$} \leq m\sub{e}c^2\sim 0.5\,$MeV the 
main source of opacity is the Compton scattering in Thomson regime. The corresponding 
optical depth is 
\begin{equation}
\tau\sub{T} \simeq \sigma\sub{T} M\sub{ej}/4\pi m\sub{p} (t v\sub{ej})^2 
      \simeq 6 M\sub{ej,1} t\sub{7}^{-2} v\sub{ej,9}^{-2} \, , 
\end{equation}
where $\sigma\sub{T}$ is the Thomson cross-section.  
Thus, the shell becomes transparent in about one year after the explosion. 
The typical energy of photons from radioactive decays is around 
$1\,$MeV and hence the ejecta opacity to these photons is similar. 
As the result, the synchrotron photons and the photons from 
radioactive decays are utilised with similar efficiency.

When $E\sub{$\nu$} \gg m\sub{e}c^2$ the scattering is in the Klein-Nishina
regime and the cross section is smaller, 
$\sigma\sub{KN}\approx 3\sigma\sub{T} m\sub{e}c^2/ 8E\sub{$\nu$}$ \citep{l80}. 
For $100\,$MeV synchrotron photons this yields 
\begin{equation}
\tau\sub{KN} \simeq 10^{-2} M\sub{ej,1} t\sub{7}^{-2} v\sub{ej,9}^{-2} \, , 
\end{equation}
and these photons begin to escape already after ten days or so. However, 
provided the spectrum of the shock accelerated electrons is a power-law 
$N(E)\propto E^{-p}$ with $p>2$, which is expected to be the case, 
the contribution of such highly energetic photons to the energy transport is 
lower.   

The IC photons with energy $E\sub{$\nu$} > (2m\sub{e} c^2)^2/E\sub{$\nu$,soft} 
\simeq 1.2\,$TeV will also interact with the soft supernova photons via the two 
photon pair-production reaction (here we used $E\sub{$\nu$,soft}=1\,$eV). 
The corresponding opacity can be estimated as 
\begin{equation}
   \tau_{\gamma\gamma} \simeq
    \frac{\sigma\sub{T}}{5} \frac{L\sub{soft}}{4\pi(v\sub{ej} t) c E\sub{soft}}
    \simeq 2 L\sub{soft,41} v\sub{ej,9}^{-1} t\sub{7}^{-1} \, .
\end{equation}
Thus, we expect the ejecta to become transparent to the IC emission 
soon after the end of the plateau phase.

The above calculations show that during the plateau phase the high 
energy emission from the termination shock of the magnetar wind is deposited 
in the ejected. However this additional heating has little effect on the 
supernova luminosity. Indeed, the increase of the luminosity due to this 
heating can be estimated as 
\begin{equation}
   \Delta L \simeq \frac{E\sub{rot}(t) t}{t\sub{d}^2} \simeq  
    \frac{L\sub{w}(t)t^2}{t\sub{d}^2} = L\sub{w}(t\sub{d}) \ll L.   
\end{equation}
For $t > t\sub{d}$ the energy deposition rate does no longer scale as $t^{-2}$ 
\citep[c.f.][]{w10} as the ejecta becomes transparent to the high energy emission.
For example, if the synchrotron emission is the main channel of heating  
the rate scales as $L\sub{w}\tau \propto t^{-4}$ when $\tau\ll 1$, 
and the rate due to IC emission is likely to decline even faster. 

Out of the well studied supernova known to the authors none fits the above 
description.  The closest example is SN1994W \citep{cv94,scl98}. This is one of 
the brightest type-II supernovae, with the peak luminosity $\sim 10^{43}\,$erg/s. 
After around 120 days its luminosity drops dramatically down to $\sim 10^{41}\,$erg/s. 
The tail of its light curve is very steep showing very small mass of ejected 
$^{56}$Ni \citep{scl98}. In fact the data can be approximated by $L\propto t^{-4.6}$. 
All these properties agree nicely with our expectations. On the negative side, 
the spectral data indicate the lower ejecta speed and the presence of a very slow 
component with $v\sim 1000\,$km/s. This supernova is classified as type-IIn and 
seems to be explained very well in the model involving a collision between the 
SN ejecta and a massive circumstellar shell ejected few years earlier in giant stellar 
eruption \citep{cbc04,wbh07,slf07,scl08}.

The power of escaped high energy emission can be estimated as 
$L_{\gamma} \simeq L\sub{w} e^{-\tau}$. As usual in such cases, 
it will peak when $\tau\sim 1$, which may occur within the first 
year after the explosion. For a source at the distance $d=10\,d\sub{1}\,$Mpc, 
the corresponding total flux will be   
\begin{equation}
     F\sups{peak} \simeq 3\by10^{-12}
     B\sub{NS,15}^{-2}
     t\sub{p,7}^{-2}
     d\sub{1}^{-2}
     \,\frac{\mbox{erg}}{\mbox{cm}^2\,\mbox{s}}, 
\label{f-syn}
\end{equation}
where $t\sub{p}$ is the time when the peak is reached. For the synchrotron 
component the spectral energy distribution is expected to peak around 
$E\sub{$\nu$}=100\,$keV (see Eq.\ref{e-peak}).  

\subsection{Remnant}

What is left behind after the explosion depends on where inside the RSG it occurs.
As we have demonstrated, this strongly depends on the mean specific angular momentum 
of gas captured via the Bondi-Hoyle mechanism. 
If it is as high as proposed in the early investigations \citep{is75,sl76} 
then the NS will accrete only at the Eddington rate until it settles into the 
core of its companion. What would happens after this is not clear. The current view 
is that the neutrino cooling will be sufficiently high for the accretion to 
proceed at the Bondi rate and that the NS collapses into a black hole 
\citep{zin72}. However, this conclusion is based on spherically symmetric models.
The high rotation rate developed in the core of the primary during the in-spiral
means that instead of directly accreting  
onto the NS the core will form a torus, whose neutrino cooling may not 
be very effective, in which case such a configuration can be relatively long-lived. 
Moreover, the mass accretion will be accompanied by efficient 
recycling of the NS, so one can still expect the NS to turn into a millisecond 
magnetar before its mass becomes high enough for gravitational collapse. 
Even if the magnetar-driven explosion fails in the 
dense environment of the core,  and the NS eventually collapses into a black 
hole, this will be a rapidly rotating black hole accompanied by a massive 
accretion disk -- the configuration characteristic of the collapsar model for 
GRB. However, in contrast to the GRB progenitors the envelope of RSG star is not 
sufficiently compact for the relativistic jet to penetrate it during the 
typical life-time of the GRB central engine. Instead, the jet energy 
will be deposited in the RSG envelope and most likely result in a 
supernova explosion, leaving behind a black hole remnant. Interestingly, 
the mass of the presupernova could be below the limit for black hole 
formation in the normal course of its evolution as a solitary star.

If the angular momentum is much lower, as suggested by \citet{dp80} then 
the explosion occurs when the NS is still inside the RSG envelope then the 
remnant is likely to be a close binary consisting of a WR star and a 
magnetar\footnote{The 2D numerical simulations by \citet{fbh96} have indicated the 
possibility of a neutrino-driven explosion during the CE phase, also leaving behind a
close binary system.}. In order to show this, we first note that the pressure inside the core  
of RSG at the He-burning phase is $p\sub{c}\sim 10^{19}\,\mbox{erg}\,\mbox{cm}^{-3}$ 
and much higher at the presupernova phase \citep{whw02}. This has to be compared 
with the pressure inside the high entropy bubble, created by the magnetar 
\begin{equation} 
p\sub{b} \simeq \frac{L\sub{w}t}{4\pi R\sub{b}^3} 
\label{pbub}
\end{equation}
where $R\sub{b} = v\sub{b}t$ is the bubble radius, $v\sub{b}$  is its expansion speed, 
and $t$ is the time since the explosion. We are interested in its value at the time 
when the bubble radius is comparable with the separation between the NS and 
the RSG core.  
Taking $R\sub{p}=10^{10}R\sub{b,10}\,$cm and $t=R\sub{b}/v\sub{b}$ we 
find that 
\begin{equation}
p\sub{b} \simeq 10^{19} L\sub{w,49} v\sub{b,9}^{-1} R\sub{b,10}^{-2} 
   \,\frac{\mbox{erg}}{\mbox{cm}^{3}}\,  . 
\label{pbub1}
\end{equation} 
If the explosion occurs while the NS is still well inside the RSG envelope 
then $R\sub{b,10}\gg 1$ and $P\sub{b}\ll P\sub{c}$. Thus, the magnetar wind 
cannot destroy the core. Another point is whether the gravitational coupling 
between the NS and the core is strong enough for the binary to survive the   
high mass loss during the explosion. Such a destruction is unlikely because 
for a sufficiently small separation at the time of the explosion most of the 
mass loss will come from the envelope, which has no effect on the gravitational 
attraction between the NS and the core.     

The magnetar wind will interact with the WR wind and will terminate when 
the radius of magnetar magnetosphere drops below the light cylinder radius. The 
magnetospheric radius is determined by the pressure balance
\begin{equation}
  \frac{B\sub{NS}^2}{8\pi}\fracb{R\sub{m}}{R\sub{NS}}^{-6} = 
  \frac{\dot{M}\sub{WR} v\sub{WR}}{4\pi a^2},
\label{p-bal}
\end{equation}
which yields 
\begin{equation}
    R\sub{m} \simeq 6\by10^9\,  
       B\sub{NS,15}^{1/3}
       \dot{M}\sub{WR,-5}^{-1/6}
       v\sub{WR,8}^{-1/6}
       a\sub{11}^{1/3} 
       \,\mbox{cm}\, .          
\label{r-mag}
\end{equation}
Using Eq.\ref{omega} to find the light cylinder radius, we estimate the termination 
time of the magnetar wind as 
\begin{equation}
   t\sub{w} \simeq 25\, 
      B\sub{NS,15}^{-4/3}
      \dot{M}\sub{WR,-5}^{-1/3} 
      v\sub{WR,8}^{-1/3}
      a\sub{11}^{2/3}
       \,\mbox{yr}\, .
\label{t-wind}
\end{equation}
The Bondi-Hoyle accretion radius of the magnetar in the WR wind 
\begin{equation}
   R\sub{A} = \frac{2GM}{v\sub{WR}^2} 
      \simeq 4\by10^{10}\, v\sub{WR,8}^{-2} 
       \,\mbox{cm}\,    
\end{equation}
is above the magnetospheric radius, indicating that the magnetar may begin 
to accrete when its co-rotation radius 
\begin{equation}
   R\sub{cor} = \fracb{GM}{\Omega^2}^{1/3} 
  \simeq 5\by10^9\, 
        B\sub{NS,15}^{2/3}
        t\sub{13}^{1/3}
       \,\mbox{cm}
\end{equation}
exceeds $R\sub{m}$.  Ignoring the decay of magnetar's magnetic field, we find that 
this will occur  at 
\begin{equation}
  t \simeq 7\by10^5\,
        B\sub{NS,15}^{-1}
      \dot{M}\sub{WR,-5}^{-1/2}
      v\sub{WR,8}^{-1/2}
      a\sub{11}
       \,\mbox{yr}\, .
\end{equation}
In fact, this time exceeds the expected decay time of the magnetar 
magnetic field 
\begin{equation}
  t\sub{dec} \simeq 10^4 B\sub{15}^{-2} 
       \,\mbox{yr}\, 
\end{equation}
\citep{hk98}.  
However, the relatively weak dependence of $R\sub{m}$ on $B\sub{NS}$ indicates
that the effect of magnetic field decay is minor and for the whole duration of 
the WR phase of the companion the magnetar will not be accreting. 
Instead, and will exhibit the behaviour typical for Soft Gamma Ray Repeaters or 
Anomalous X-Ray Pulsars \citep{TD96}.  

Normally, a close high-mass binary system can produce at 
most two supernovae. However, our results suggest that in some cases this 
number can be increased to three. 
Indeed, in our model the first explosion creates the NS itself. 
When this NS  spirals into its companion the second, now off-center explosion 
occurs.  Finally, when the core of the WR star eventually collapsed, there 
is another explosion.  
Moreover, the very close separation between the magnetar and WR star 
makes this binary system a promising GRB progenitor \citep[e.g.][]{BK10}.
Indeed, the WR star is very compact so the relativistic GRB jets can break out 
in a reasonably short time and the spin-orbital interaction ensures that the WR star 
is rapidly rotating. These are the two most important properties which a 
long GRB progenitor must have.

\section{Conclusions}
\label{conclusions}

The main aim of this study was to investigate whether during the common 
envelope phase of a close binary, involving RSG and NS, 
the NS can spin up to a millisecond period and generate magnetar-strength magnetic field.  
If possible, this would result in magnetically driven stellar explosion, releasing 
up to $10^{52}\,$erg of magnetar's rotational energy inside the RSG. 

It turns out that the outcome is very sensitive to the specific angular momentum 
of the gas gravitationally captured by the NS during the in-spiral. Should it be 
as high as suggested in the early papers by \citet{is75} and \citet{sl76}, the accretion 
rate is low and neither mass nor spin of the NS will increase significantly until 
the NS settles into the center of RSG. This is because the accretion disk  
forms too far from the NS, its temperature remains rather low, and it is
unable to cool effectively via neutrino emission, which is very sensitive to 
temperature. As the result, the accretion shock is pushed 
too far, beyond the sonic point of the Bondi-type flow, thus preventing the accretion 
from reaching the Bondi-Hoyle rate \citep{ch96}. The CE can either be ejected, leaving 
behind a close NS-WR binary or survive, leading to a merger of the NS with the 
RSG core \citep{Taams00}. It is not clear what exactly would 
occur following the merger, as we cannot treat this case using our approach. 
The common belief is that the NS will begin to accrete at the Bondi rate 
and collapse into a black hole. However, this conclusion is based on models with 
spherical symmetry, whereas the strong rotation developed by the system during the 
merger makes the assumption of spherical symmetry unsuitable. It may still be possible 
that the NS spins up to a millisecond period and becomes a magnetar 
before its mass reaches the limit of gravitational collapse. 

If, on the contrary, the angular momentum is much lower, more in line with 
the analysis of \citet{dp80}, the accretion can begin to proceed at the Bondi-Hoyle 
rate when the NS is still inside the common envelope. Further investigation of 
the Bondi-Hoyle accretion via 3D numerical simulations are required to clarify 
this issue.   

Once the NS becomes a millisecond magnetar, its emerging magnetosphere interacts 
with the accretion flow. If this occurs while the NS is still inside the common 
envelope and the magnetic field is as strong as $\sim10^{15}$G, the interaction is 
most likely to begin in the propeller regime and then quickly proceed to the 
ejector regime, given the small radius of the light cylinder. Strong magnetar wind 
blows away the common envelope and drives an explosion whose energy is likely to 
exceed the standard $10^{51}$erg of normal supernovae. However, 
the core of the companion survives and the explosion leaves behind a very close 
binary, consisting of a magnetar and a WR star. In spite of the very close 
separation and the very strong wind from the WR star, the NS is shielded from 
the wind by its magnetosphere and prevented from accreting.  Later on, when the WR explodes 
this will be a {\it third} explosion produced in the system. Moreover, the rapid rotation 
and the compactness of this WR star show that this explosion can be accompanied 
by a gamma-ray burst \citep{BK10}.    

If the magnetar forms only after the merger with the core then a higher magnetic 
field, $\sim10^{16}$G, is required to explode the star. In this case, the remnant is 
a solitary magnetar. If, however, the magnetar-driven explosion fails and the NS collapses 
into a black hole, this will be a rapidly rotating black with a massive accretion 
disk. Although this is exactly the configuration proposed in the collapsar 
model of gamma-ray bursts, the relativistic jets will not be able to escape from the 
extended envelope of RSG and produce such a burst. Instead, they will deposit their 
energy inside the envelope and drive a type-II supernova.         

Given the very high rotational energy of a millisecond NS the supernova is 
expected to be very bright at the plateau phase and show very broad spectral 
lines. However, only a small amount of $^{56}$Ni is expected to be produced in
the explosion due to the relatively low densities, in comparison to those reached 
during the normal core-collapse explosions. By the end of the plateau phase 
the power of the magnetar wind is also significantly reduced and the supernova 
brightness is expected to drop sharply. The combination of the rapidly declining 
power of magnetar wind and the increasing transparency of the supernova ejecta to 
the high energy emission from the wind termination shock result in 
steeper than normal light-curve tails.  

The unique property of supernovae produced in this way is the high energy 
synchrotron and inverse Compton emission from the magnetar wind nebula. 
Unfortunately, the supernova ejecta does not become transparent to this emission 
until a hundred days after the explosion, by which time the NS rotation 
is already very slow and its wind is no longer that powerful. 
The flux of gamma-ray emission is expected to be rather low and difficult to 
observable, unless the explosion occurs in the Local Group of galaxies.

\section{Acknowledgments}
We are thankful to V.Bosch-Ramon, N.Ikhsanov, and Y.Levin for fruitful 
discussions, as well as to A.Heger for providing models of RSGs. 
The calculations were fulfilled at cluster of Moscow State University "Chebyshev". 
This research was funded by STFC under the Rolling Programme of 
Astrophysical Research at Leeds University.

\end{document}